\documentclass[
    reprint, twocolumn,
    aps,prl,10pt,amsmath,amssymb,
    longbibliography,superscriptaddress,
    showpacs,preprintnumbers,
]{revtex4-2}

\usepackage{xr}
\makeatletter
\newcommand*{\addFileDependency}[1]{
  \typeout{(#1)}
  \@addtofilelist{#1}
  \IfFileExists{#1}{}{\typeout{No file #1.}}
}
\makeatother

\newcommand*{\myexternaldocument}[1]{
    \externaldocument{#1}
    \addFileDependency{#1.tex}
    \addFileDependency{#1.aux}
}

\myexternaldocument{supp}

\usepackage[colorlinks=true, linkcolor=blue, citecolor=blue, urlcolor=blue]{hyperref}
\usepackage{oubraces}
\usepackage{bm,bbm,mathbbol}
\usepackage{physics}
\usepackage{mathtools}
\usepackage{amsmath, amsthm, amssymb, amsfonts}
\usepackage{bbold}
\usepackage{esvect}
\usepackage[dvipsnames]{xcolor}
\usepackage{epsfig}
\usepackage{array}
\usepackage{multirow}
\usepackage{graphicx}
\usepackage[normalem]{ulem}
\usepackage{float}
\usepackage[section]{placeins}
\usepackage{afterpage}
\usepackage{xifthen}

\usepackage[caption=false]{subfig}
\usepackage{multirow}
\usepackage{placeins}
\usepackage{cancel}

\usepackage{tabularx}   
\newcolumntype{Y}{>{\centering\arraybackslash}X}  

\newcommand{\mb}[1]{{\bm{\mathrm{#1}}}}  

\newcommand{\veps}[0]{\varepsilon}

\newcommand{\one}[0]{\mathbbm{1}}

\newcommand{\ccdots}{{{\cdot}{\cdot}{\cdot}}} 
\newcommand{\stkout}[1]{\ifmmode\text{\sout{\ensuremath{#1}}}\else\sout{#1}\fi}

\newcommand{\nk}[0]{{n\mathbf{k}}}

\newcommand{\mkq}[0]{{m\mathbf{k+q}}}
\newcommand{\qnu}[0]{{\mathbf{q}\nu}}

\newcommand{\Eq}[1]{Eq.~\eqref{#1}}

\newcommand{\Sec}[1]{Sec.~\ref{#1}}

\newcommand{\Fig}[1]{Fig.~\ref{#1}}

\newcommand{\plph}[0]{{\mathrm{pl\text{-}ph}}}
\newcommand{\qz}[0]{{\mb{q}0}}
\newcommand{\qw}[0]{{\mb{q}, \omega}}
\newcommand{\qmu}[0]{{\mb{q}\mu}}
\newcommand{\qa}[0]{{\mb{q}\alpha}}

\newcommand{\qb}[0]{{\mb{q}\beta}}
\newcommand{\aap}[0]{{\alpha\alpha'}}
\newcommand{\Nph}[0]{{N_{\rm ph}}}

\usepackage{orcidlink}

\begin{document}

\title{Plasmon-phonon hybridization in doped semiconductors from first principles}

\author{Jae-Mo Lihm\,\orcidlink{0000-0003-0900-0405}}
\email{jaemo.lihm@gmail.com}
\author{Cheol-Hwan Park\,\orcidlink{0000-0003-1584-6896}}
\email{cheolhwan@snu.ac.kr}
\affiliation{Department of Physics and Astronomy, Seoul National University, Seoul 08826, Korea,}
\affiliation{Center for Correlated Electron Systems, Institute for Basic Science, Seoul 08826, Korea,}
\affiliation{Center for Theoretical Physics, Seoul National University, Seoul 08826, Korea}

\date{\today}

\begin{abstract}
Although plasmons and phonons are the collective excitations that govern the low-energy physics of doped semiconductors, their nonadiabatic hybridization and mutual screening have not been studied from first principles.
We achieve this goal by transforming the Dyson equation to the frequency-independent dynamical matrix of an equivalent damped oscillator.
Calculations on doped GaAs and TiO$_2$ agree well with available Raman data and await immediate experimental confirmation from infrared, neutron, electron-energy-loss, and angle-resolved photoemission spectroscopies.
\end{abstract}

\footnotetext[1]{See Supplemental Material [URL will be inserted by publisher]
for the derivation of the plasmon-phonon Green function and electron-plasmon-phonon coupling and computational details, which includes Refs.~\cite{Verdi2015Frohlich,Sjakste2015Frohlich,McWeenyBook,Bauernschmitt1996TDDFT,Filip2021,MahanBook,StefanucciBook,Thygesen2007GW,Ness2011GW,2020PonceReview,2017GiannozziQE,2013HamannONCVPSP,2018VanSettenPseudoDojo,Perdew1992LDA,Horn1972TiO2,2020PizziWannier90,Giustino2007,2016PonceEPW,1997MarzariMLWF,2012MarzariMLWF,2017BezansonJulia,2021PonceMobility,2001SouzaMLWF}.}
\newcommand{\citeSupp}[0]{Note1}

\maketitle

\begin{table}[b]
\begin{tabular}{ccccc}
\hline
Regime         & Energies                                  & \begin{tabular}[c]{@{}c@{}}LO-TO\\ splitting\end{tabular} & \begin{tabular}[c]{@{}c@{}}Long-range\\ electric field\end{tabular} &  \\ \hline
Anti-adiabatic & $\omega_{\rm pl} \ll \omega_{\rm ph}$     & O$\hphantom{^*}$     & phonon &  \\ \hline
Resonant       & $\omega_{\rm pl} \approx \omega_{\rm ph}$ & O$^*$ & pl-ph hybrid &  \\ \hline
Adiabatic      & $\omega_{\rm pl} \gg \omega_{\rm ph}$     & X$\hphantom{^*}$ & plasmon &  \\ \hline
\end{tabular}
\caption{
    Three regimes of plasmon-phonon (pl-ph) hybridization.
    $\omega_{\rm pl}$ and $\omega_{\rm ph}$ denote the plasmon and phonon energy scales, respectively.
    (O$^*$: The LO phonons form a plasmon-phonon hybrid, making the concept of LO-TO splitting unclear.)
}
\label{tab:summary}
\end{table}

Doped semiconductors play the most essential role in modern electronics.
An important property of doped semiconductors is the presence of low-energy collective charge excitations, plasmons.
Plasmons in doped semiconductors strongly couple to other low-energy excitations, particularly phonons.
This coupling determines the low-energy spectroscopic properties of doped semiconductors.
In addition, electron-phonon and electron-plasmon interactions are the two major contributions to electron scattering at room temperature.
Therefore, understanding and modeling the interplay of plasmons and phonons and their coupling to electrons is crucial for studying doped semiconductors.

Plasmon-phonon coupling can be divided into three regimes in order of increasing doping: anti-adiabatic, resonant, and adiabatic~\cite{2017Verdi,Riley2018Polaron}.
The change in the plasmon energy relative to the phonon energy governs the crossover from insulators to metals and affects many properties, such as the splitting between longitudinal optical (LO) and transverse optical (TO) phonons and the presence of long-range electric fields (Tab.~\ref{tab:summary}).
At low doping, the phonon energy is higher; hence, the plasmons do not screen the phonons, and both the LO-TO splitting and phonon-induced long-range electric fields are present, just as in undoped systems.
At intermediate doping, where the plasmon energies are comparable to the phonon energies, the two modes hybridize strongly~\cite{Yokota1961PlPh,Varga1965PlPh,Singwi1966PlPh,Cochran1966PlPh}.
This hybridization leads to a level anticrossing behavior that was confirmed by Raman experiments on GaAs~\cite{Mooradian1966PlPh,Mooradian1967PlPh}.
At even higher doping, the plasmon energy exceeds the LO phonon energy, and the plasmons fully screen the electric field generated by the phonons.
The LO-TO splitting and the electron-phonon coupling are then strongly suppressed.

\begin{figure}[tb]
\centering
\includegraphics[width=0.99\linewidth]{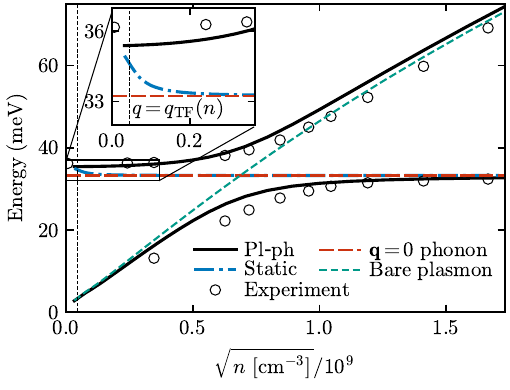}
\caption{
    Doping-dependent energy of the plasmon-phonon (pl-ph) hybrids of $n$-doped GaAs.
    The solid black (dash-dotted blue) curves show the result of a nonadiabatic plasmon-phonon (static) calculation.
    The circles show the experimental measurements of Ref.~\cite{Mooradian1967PlPh}.
    The horizontal dashed red line shows the optical phonon energy at $\mb{q}=0$, which corresponds to the TO phonon energy in the case of GaAs.
    The green dashed line shows the bare plasmon energy.
    We use the experimental wavevector $q = 8\times10^{-4}~\mathrm{\AA}^{-1}$ in the [110] direction.
    The vertical dotted line denotes the density $n=2\times10^{15}~\mathrm{cm}^{-3}$ where the Thomas-Fermi wavevector $q_{\rm TF}$ is equal to the experimental wavevector.
    All calculations in this paper are performed for $T=300~\mathrm{K}$.
}
\label{fig:dispersion_GaAs}
\end{figure}

A major hurdle in studying plasmon-phonon coupling in real materials has been the lack of an efficient first-principles methodology.
In the anti-adiabatic and resonant regime, one needs a nonadiabatic description of phonons~\cite{Saitta2008,2010CalandraPhonon,Novko2018,Novko2020} that goes beyond the Born--Oppenheimer approximation.
In particular, one needs to capture the full frequency dependence of the dielectric function and the phonon self-energy~\cite{2017GiustinoRMP}.
Due to the high computational cost of such calculations, few studies considered the nonadiabatic coupling~\cite{Falter1993,Bauer2009Plasmon,Krsnik2022PlPh,Kengle2023RPA}, relying on simplified models of the electron or phonon dispersion.
Previous first-principles studies on doped semiconductors, namely Refs.~\cite{2017Verdi,Caruso2018}, have neglected any feedback from phonons to plasmons and used the electron gas model to compute the free-carrier screening.
References~\cite{2022MachedaDoping,Macheda2023Doping} represent a different line of work, which treats doping using the static density functional perturbation theory (DFPT) and thus cannot reproduce any nonadiabatic effects.
Another important open problem is to understand how the plasmon-phonon modes couple to the electrons.
While there have been separate first-principles calculations on the electron-phonon and electron-plasmon couplings~\cite{Caruso2016Plasmon,Caruso2018}, a unified first-principles theory of the electron-plasmon-phonon coupling has not been developed.
Developing a first-principles method for calculating the plasmon-phonon hybridization in all three regimes remains an open challenge in modeling doped semiconductors~\cite{2017GiustinoRMP,Caruso2020Review}.

\begin{figure}[tb]
\centering
\includegraphics[width=0.99\linewidth]{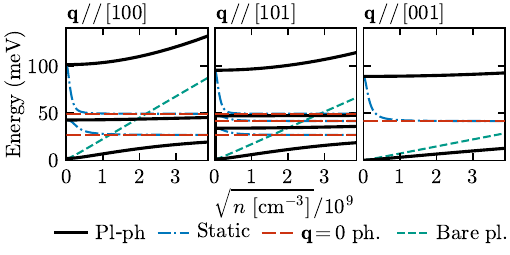}
\caption{
    Doping- and wavevector-direction-dependent energy of the plasmon-phonon hybrids of $n$-doped anatase TiO$_2$ for wavevectors of magnitude $q = 2\times10^{-3}~\mathrm{\AA}^{-1}$.
    Only the hybridized modes are plotted; the other, pure TO or infrared-inactive phonon modes do not hybridize with plasmons.
}
\label{fig:dispersion_TiO2}
\end{figure}

\begin{figure*}[tb]
\centering
\includegraphics[width=0.99\linewidth]{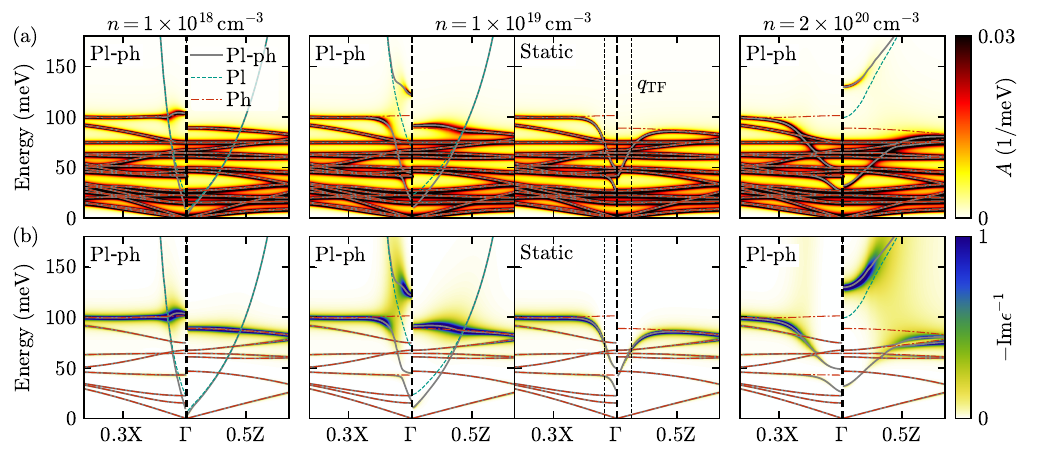}
\caption{
    (a) Phonon spectral function and (b) inverse dielectric function of $n$-doped anatase TiO$_2$.
    We add an artificial broadening of 0.5~meV to the bare phonon Green functions for visualization.
    The third column shows the results obtained with the static approximation, while the others show the results of nonadiabatic plasmon-phonon calculations.
    The dotted vertical lines in the third column denote the Thomas-Fermi wavevector, the wavevector where the static dielectric function equals twice $\epsilon^\infty_{\hat{q}}$.
    In (b), we plot only the modes contributing to the dielectric function.
}
\label{fig:spectral}
\end{figure*}

\begin{figure}[tb]
\centering
\includegraphics[width=0.99\linewidth]{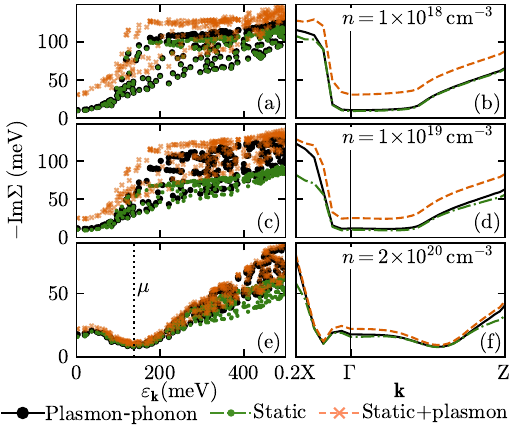}
\caption{
    (a) Imaginary electron self-energy of $n$-doped anatase TiO$_2$ at $n=1\times10^{18}~\mathrm{cm}^{-3}$ as a function of electron energy.
    $\veps_\mb{k} = 0$ is the conduction band minimum energy.
    (b) The same quantity as in (a) as a function of the electron wavevector.
    (c-f) Same as in (a, b) but for $n=1\times10^{19}~\mathrm{cm}^{-3}$ [(c,d)] and $n=2\times10^{20}~\mathrm{cm}^{-3}$ [(e,f)].
    In (e), the vertical dotted line indicates the chemical potential $\mu$, while in (a) and (c) $\mu$ is below the conduction band minimum.
}
\label{fig:selfen}
\end{figure}

In this study, we develop a fully first-principles description of the plasmon-phonon hybridization in doped semiconductors.
We transform the \textit{nonadiabatic} problem of computing the phonon Green function into an \textit{adiabatic} problem of diagonalizing a plasmon-phonon dynamical matrix, significantly reducing the computational cost.
We first validate our new method by quantitatively reproducing {\it without any free parameters} the dispersion of plasmon-phonon hybrids in GaAs~\cite{Mooradian1967PlPh} (\Fig{fig:dispersion_GaAs}).
We then study the complicated wavevector-direction-dependent plasmon-phonon coupling of anatase TiO$_2$, presenting many predictions that await immediate experimental confirmation by infrared (\Fig{fig:dispersion_TiO2}), neutron scattering [Fig.~\ref{fig:spectral}(a)], electron energy loss [Fig.~\ref{fig:spectral}(b)], and angle-resolved photoemission (\Fig{fig:selfen}) spectroscopies.
Our first-principles method can also be widely used to study polarons, transport, and superconductivity of doped semiconductors.


To model plasmon-phonon hybridization, we first perform a static DFPT calculation of the \textit{undoped} system.
We then treat the doping using the rigid-band approximation and incorporate the screening effects in the random phase approximation.
The doped carriers induce a phonon self-energy~\cite{2010CalandraPhonon,Nomura2015cDFPT,2023BergesPhonon,Marini2023}
\begin{equation} \label{eq:ph_selfen}
    \Pi_{\mu\nu}(\qw)
    = (g_\qmu^{\rm lr})^* \, \delta \chi(\qw) \, g_\qnu^{\rm lr},
\end{equation}
where $\mb{q}$ is the phonon wavevector, $\omega$ the frequency, $\mu, \nu = 1, \ccdots, \Nph$ the phonon mode indices, $\delta \chi(\qw)$ the interacting susceptibility of the doped carriers~[\Eq{eq:epsilon}],
and $g_\qmu^{\rm lr}$ the long-range electron-phonon vertex~[\Eq{eq:eph_lr}].
This self-energy is nonadiabatic, i.e., frequency dependent, and thus leads to a nontrivial renormalization that gives rise to additional peaks in the phonon spectral function.

When calculating the interacting electronic susceptibility in the random phase approximation, we neglect local field effects, as has been validated for doped semiconductors~\cite{Liang2015Plasmon,2022MachedaDoping}.
We model the frequency dependence of the susceptibility using a plasmon-pole model~\cite{Hybersten1986GW}:
\begin{equation} \label{eq:plasmon_pole_model}
    \delta\chi(\qw)
    = \frac{1}{U_\mb{q}} \frac{\Omega_\qz^2}{(\omega + i\gamma_\qz)^2 - \omega_\qz^2},
\end{equation}
where $U_\mb{q} = 4\pi e^2/(V q^2\epsilon^\infty_{\hat{q}})$ with $\epsilon^\infty_{\hat{q}} = \hat{q} \cdot \mb{\epsilon}^\infty \!\cdot \hat{q}$ is the macroscopic Coulomb interaction screened by the electronic dielectric tensor $\mb{\epsilon}^\infty$ of the undoped system and $V$ the unit cell volume.
The model has three parameters, $\Omega_\qz$, $\omega_\qz$, and $\gamma_\qz$: the plasmon strength, frequency, and linewidth.
The plasmon linewidth is non-zero in the electron-hole continuum and captures the decay of plasmons into electron-hole pairs.
Our method can be straightforwardly generalized to treat more complex frequency dependences using the multipole approach~\cite{Leon2021Multipole} (see Sec.~S1.E for details).

With the bare phonon Green function
\begin{equation} \label{eq:ph_bare_D}
    D^{0}_{\mu\nu}(\qw) = \delta_{\mu\nu} \frac{1}{(\omega + i 0^+)^2 - \omega_\qnu^2},
\end{equation}
where $\omega_\qnu$ is the phonon frequency of the undoped system, the full Green function $D$ is computed by solving the Dyson equation
\begin{equation} \label{eq:ph_Dyson}
    D^{-1}_{\mu\nu}(\qw)
    = (D^{0})^{-1}_{\mu\nu}(\qw) - \Pi_{\mu\nu}(\qw).
\end{equation}
Solving \Eq{eq:ph_Dyson} at each $\omega$ inflates the computational cost, making this calculation unsuitable for a first-principles application dealing with a large number of wavevectors or materials with a complex unit cell.

Remarkably, it was found that one can exactly transform the problem of solving a \textit{frequency-dependent} fermionic Dyson equation into a simpler problem of diagonalizing a \textit{frequency-independent} matrix by writing the self-energy as a sum over poles, in the context of dynamical mean-field theory~\cite{Savrasov2006AIM} and $GW$~\cite{Chiarotti2022GW} calculations.
In a different context, semiclassical studies of phonon-polaritons~\cite{Bonini2022Cavity} and spin-lattice coupling~\cite{Ren2024} have developed dynamical matrices that describe the hybridization of phonons with other bosonic modes.
We combine these fascinating ideas and apply them to bosonic plasmon-phonon hybridization, thereby developing a method of solving the phonon Dyson equation via a \textit{frequency-independent}, $(N_{\rm ph}+1)$-dimensional plasmon-phonon dynamical matrix.
In addition, we find that one can incorporate finite bosonic linewidths, which are essential for our study, by diagonalizing a $2(N_{\rm ph}+1)$-dimensional dynamical matrix for a damped oscillator.

If the plasmon linewidth is zero ($\gamma_\qz = 0^+$), the phonon Green function has $\Nph+1$ poles whose energy is the square root of the eigenvalues of an $(\Nph+1)$-dimensional plasmon-phonon dynamical matrix
\begin{equation} \label{eq:ext_dynamical_matrix}
    \widetilde{C}_{\mb{q}}
    = \begin{pmatrix}
        \omega_{\mb{q}0}^2 & c_{\mb{q}1} & c_{\mb{q}2} & \cdots
        \\
        c_{\mb{q}1}^* & \omega_{\mb{q}1}^2 & 0 & \cdots
        \\
        c_{\mb{q}2}^* & 0 & \omega_{\mb{q}2}^2 & \cdots
        \\
        \vdots & \vdots & \vdots & \ddots
    \end{pmatrix},
\end{equation}
where the plasmon and phonon frequencies are on the diagonals and the plasmon-phonon coupling amplitudes
\begin{equation}
    c_\qnu = \Omega_\qz\, g^{\rm lr}_\qnu / \sqrt{U_\mb{q}}
\end{equation}
on the wings.
We denote objects in the $(\Nph+1)$-dimensional plasmon-phonon basis by a tilde.
They have indices $\alpha = 0, \cdots, \Nph$, where 0 corresponds to the plasmon, and the others to the phonons.

We diagonalize $\widetilde{C}_\mb{q}$ with orthonormal eigenvectors as
\begin{equation}
    \sum_{\alpha'=0}^{\Nph} \widetilde{C}_{\mb{q} \alpha\alpha'} \widetilde{V}'_{\mb{q} \alpha' \beta}
    = \widetilde{\omega}_\qb^2 \widetilde{V}'_{\mb{q} \alpha \beta}.
\end{equation}
The eigenvalue $\widetilde{\omega}_\qb$ is the energy of a plasmon-phonon hybrid mode, and the eigenvector $\widetilde{V}'_{\mb{q} \alpha \beta}$ encodes the weight of the plasmon ($\alpha=0$) and phonons ($\alpha=1,\cdots,\Nph$) of the hybrid mode.
The phonon Green function can be written as a sum over these hybrid modes:
\begin{equation}
    D_{\mu\nu}(\qw)
    = \sum_{\beta=0}^{\Nph} \frac{\widetilde{V}'_{\mb{q}\mu \beta} \widetilde{V}_{\mb{q}\nu\beta}^{\prime*}}{(\omega + i0^+)^2 - \widetilde{\omega}_\qb^2}.
\end{equation}
The dielectric function can be calculated similarly~[\Eq{eq:dielec_epsil_nobroad}]~\cite{\citeSupp}.

One can derive analogous results for non-zero plasmon linewidth $\gamma_\qz \neq 0$ by solving a damped oscillator problem, whose dynamical matrix in the generalized position-velocity coordinate basis is given by
\begin{equation}
    \begin{pmatrix}
        -i \widetilde{\Gamma}_{\mb{q}} & \one \\
        \widetilde{C}_{\mb{q}} & - i\widetilde{\Gamma}_{\mb{q}}
    \end{pmatrix}\,,
\label{eq:2N_plus_2_matrix}
\end{equation}
where $\widetilde{\Gamma}_{\mb{q}\aap} = \delta_{\aap}\, \gamma_\qa$ and $\one$ is an $(\Nph+1)$-dimensional identity matrix.
Now, the plasmon-phonon hybrids have complex eigenvalues whose real and imaginary parts correspond to the energy and linewidth of the hybrid mode, respectively.
See \Sec{sec:supp_deriv} for the details of the derivation of Eqs.~(\ref{eq:ext_dynamical_matrix}-\ref{eq:2N_plus_2_matrix})~\cite{\citeSupp}.


We first apply our plasmon-phonon method to GaAs, a prototypical material for studying plasmon-phonon hybridization~\cite{Mooradian1966PlPh,Mooradian1967PlPh}.
Figure~\ref{fig:dispersion_GaAs} shows the calculated doping dependence of the plasmon-phonon hybrid energies at 300~K, compared to the measured Raman data~\cite{Mooradian1967PlPh}.
We find good quantitative agreement between the experiment and theory without any free parameters.
The static approximation completely fails to capture the plasmon-phonon hybridization, leading to a vanishing LO-TO splitting at $n >2\times10^{15}~\mathrm{cm}^{-3}$, the density over which the Thomas-Fermi wavevector exceeds the experimental momentum transfer.
Moreover, for an infinitesimal $q$, the static approximation gives vanishing LO-TO splitting at all doping levels, while the dispersion obtained from the nonadiabatic calculation remains invariant.

We now study plasmon-phonon coupling in anatase TiO$_2$.
Anatase is a stable, naturally abundant, and nontoxic material widely used in photocatalytic, photovoltaic, and electronic applications~\cite{Toyosaki2004TiO2,Furubayashi2005TiO2,Kuang2008TiO2,Strukov2008TiO2,Varghese2009TiO2,DeAngelis2014TiO2}.
Due to its highly tunable electron doping, which is achievable up to $3.5\times10^{20}~\mathrm{cm}^{-3}$~\cite{Moser2013TiO2}, and strong electron-phonon coupling, TiO$_2$ has been a test bed for studying electron-phonon coupling in doped semiconductors, both experimentally~\cite{Moser2013TiO2,Setvin2014TiO2,Moser2015TiO2} and theoretically~\cite{2017Verdi,Caruso2018,Krsnik2022PlPh}.

Figure~\ref{fig:dispersion_TiO2} shows the energies of the plasmon-phonon hybrids for wavevectors with infinitesimal magnitudes at 300~K.
Due to the tetragonal lattice structure with a $\mathrm{D_{4h}}$ point group symmetry, one to three phonon modes couple to the plasmons, depending on the direction of the wavevector.
The remaining, purely TO or infrared-inactive phonons do not couple to the plasmons; hence, their energies are doping-independent.
The rich doping dependence and anisotropy of the hybrid modes can be confirmed by infrared spectroscopy.

Figure~\ref{fig:spectral} shows the doping-dependent phonon spectral function and the imaginary part of the inverse dielectric function, which can be measured by neutron scattering and electron energy loss spectroscopies, respectively.
Due to the highly anisotropic effective mass, the plasmon energy is anisotropic and discontinuous at $\mb{q}=\Gamma$~\cite{Ahn2021Plasmon}.
Our first-principles calculations fully reveal the anisotropic doping dependence of the plasmon-phonon hybridization.
At low doping ($n=1\times10^{18}~\mathrm{cm}^{-3}$), the in-plane and out-of-plane plasma frequencies are both below the corresponding LO frequencies, and TiO$_2$ is in the anti-adiabatic regime.
At $n=1\times10^{19}~\mathrm{cm}^{-3}$, the in-plane LO phonons and plasmons are resonant, while the out-of-plane LO phonon is still in the anti-adiabatic regime.
At $n=2\times10^{20}~\mathrm{cm}^{-3}$, the in-plane LO phonons are fully screened and are in the adiabatic regime with vanishing LO-TO splitting, while the out-of-plane phonons are in the resonant regime.
This behavior is also reflected in the inverse dielectric function, where the signal is strongest for the LO phonon, hybrid modes, and plasmons in the anti-adiabatic, resonant, and adiabatic regimes, respectively.
Moreover, we find a momentum-direction- and momentum-magnitude-dependent anticrossing between plasmons and phonons, as their bare energies approximately scale as $q^2$ and $q^0$, respectively.

Nonadiabaticity is crucial for the description of phonons in the anti-adiabatic and resonant regimes.
Unlike the results from the nonadiabatic calculations, static screening yields adiabatic phonons with vanishing LO-TO splitting, continuous spectral functions, and no plasmon-phonon hybridization.
It also overly screens the phonons and underestimates the inverse dielectric functions, especially for wavevectors smaller than the Thomas-Fermi wavevector.
These problems (see the third column in Fig.~\ref{fig:spectral}) arise from the false assumption of perfect adiabatic screening in the Born--Oppenheimer approximation.


Our new method allows one to efficiently study the coupling of plasmon-phonon hybrids to electrons without the need for explicit frequency integration.
Only minor modifications to the existing framework for electron-phonon coupling calculations are required.
For example, the electron self-energy can be calculated as the sum over the $\Nph+1$ hybrid modes, just as the sum over the phonons in electron-phonon calculations:
\begin{align} \label{eq:el_sigma}
    \mathrm{Im} \Sigma^{\plph}_\nk
    &= \mathrm{Im} \frac{1}{N_q} \sum_{m\mb{q}} \sum_{\beta=0}^{\Nph} \sum_{\pm}
    \frac{\abs{\widetilde{g}_{mn\beta}(\mb{k},\mb{q})}^2 (f^\pm_\mkq + \widetilde{n}_\qb)}{\veps_\nk - \veps_\mkq \pm \widetilde{\omega}_\qb + i 0^+}.
\end{align}
Here, $\widetilde{g}_{mn\beta}$ is the electron-hybrid coupling vertex~[\Eq{eq:eph_gtilde_nobroad}],
$N_q$ the number of sampled $q$ points, 
$f^+_\mkq$ the Fermi--Dirac occupation at the electron energy $\veps_\mkq$,
$f^-_\mkq = 1 - f^+_\mkq$,
and $\widetilde{n}_\qb = 1 / \bigl(e^{\widetilde{\omega}_\qb /k_{\rm B} T} - 1\bigr)$ the Bose--Einstein occupation of the plasmon-phonon hybrids.
See \Sec{sec:supp_eph} for details~\cite{\citeSupp}.

Figure~\ref{fig:selfen} shows the imaginary part of the electron self-energy, which can be directly measured by angle-resolved photoemission spectroscopy~\cite{Bostwick2007,Park2007,Park2009}.
The self-energy arising from statically screened phonons~\cite{2022MachedaDoping} can be significantly lower than the correct, hybrid-induced self-energy.
A possible \textit{ad hoc} correction is to add the bare-plasmon-induced self-energy~\cite{Caruso2016Plasmon,Caruso2018} to the static-phonon-induced self-energy.
This correction works well in the metallic regime.
However, it overestimates the self-energy when the electron concentration is low because
the screening of the electron-plasmon coupling by higher-energy phonons is neglected.

Our plasmon-phonon method opens the door to first-principles studies of plasmons and phonons in doped semiconductors.
Because of its simplicity and efficiency, it can be widely used for many applications: various spectroscopies, optical properties, transport, and superconductivity.
The calculated plasmon-phonon dispersion, spectral function, dielectric function, and electron self-energy can be directly compared with experiments such as infrared, Raman, neutron, electron energy loss, and angle-resolved photoemission spectroscopies.
Formation of polaronic satellites due to plasmon-phonon hybrids~\cite{Caruso2015,2017Verdi,Riley2018Polaron} and superconductivity in dilute materials~\cite{Prakash2017,Gastiasoro2020SrTiO3} are also interesting avenues for future research.

\begin{acknowledgments}
This work was supported by the Creative-Pioneering Research Program through Seoul National University, Korean NRF No-2023R1A2C1007297, and the Institute for Basic Science (No. IBSR009-D1).
Computational resources have been provided by KISTI (KSC-2022-CRE-0407).
\end{acknowledgments}

\FloatBarrier 

\makeatletter\@input{xy.tex}\makeatother
\bibliography{main}

\end{document}


\title{Supplemental Material:\\
Plasmon-phonon hybridization in doped semiconductors from first principles}

\author{Jae-Mo Lihm\,\orcidlink{0000-0003-0900-0405}}
\email{jaemo.lihm@gmail.com}
\author{Cheol-Hwan Park\,\orcidlink{0000-0003-1584-6896}}
\email{cheolhwan@snu.ac.kr}
\affiliation{Department of Physics and Astronomy, Seoul National University, Seoul 08826, Korea,}
\affiliation{Center for Correlated Electron Systems, Institute for Basic Science, Seoul 08826, Korea,}
\affiliation{Center for Theoretical Physics, Seoul National University, Seoul 08826, Korea}

\date{\today}

\maketitle


\section{Derivation of the plasmon-phonon Green function} \label{sec:supp_deriv}

In this section, we present the derivation of the plasmon-phonon dynamical matrix, which allows an efficient calculation of the phonon Green function.
First, we discuss the free-carrier dielectric function and its plasmon-pole approximation.
Then, we describe the phonon Dyson equation where the plasmon-phonon coupling induces a self-energy to the phonon.
Finally, we detail how to compute the phonon Dyson equation by solving a \textit{frequency-independent} problem in the subspace of the plasmon-phonon hybrid modes.
We set $\hbar=1$ and $4\pi\epsilon_0 = 1$, where $\epsilon_0$ is the vacuum permittivity.

\subsection{Free-carrier dielectric function} \label{eq:sec:supp_dielectric}
Using the random phase approximation (RPA) and neglecting local-field effects, which have been shown to be good approximations for doped semiconductors~\cite{Liang2015Plasmon,2022MachedaDoping}, we find that the bare susceptibility and dielectric function due to the free carriers are
\begin{equation} \label{eq:plph_chi0}
    \delta\chi_0(\qw)
    = \frac{1}{N_k} \sum_{mn\mb{k}} \frac{
        (\delta f_\mkq - \delta f_\nk) \abs{\braket{u_\mkq}{u_\nk}}^2
    }{
        \veps_\mkq - \veps_\nk + \omega + i0^+
    },
\end{equation}
\begin{equation} \label{eq:epsilon}
    \epsilon_{\rm el}^{-1}(\qw)
    = 1 + U_\mb{q} \delta \chi(\qw)
    = [1 - U_\mb{q} \delta \chi_0(\qw)]^{-1},
\end{equation}
where $N_k$ is the number of sampled $k$ points, $\delta f_\nk$ the doped carrier occupation at band $n$ and wavevector $\mb{k}$, $\veps_\nk$ the corresponding band energy, and $\ket{u_\nk}$ the periodic part of the electron wavefunction.
$\epsilon^\infty_{\hat{q}} = \hat{q} \cdot \mb{\epsilon}^\infty \!\cdot \hat{q}$ and $U_\mb{q} = 4\pi e^2/(q^2 \epsilon^\infty_{\hat{q}} V)$ are the infinite-frequency dielectric constant and screened Coulomb interaction of the undoped system, respectively, with $V$ the volume of a unit cell.
We compute $\delta\chi_0$ using Wannier interpolation by approximating $\braket{u_\mkq}{u_\nk}$ as the overlap of the eigenvectors in the Wannier basis, which is valid for $q\approx 0$~\cite{Verdi2015Frohlich,2022MachedaDoping}, where the macroscopic screening is most relevant.

Next, we represent the frequency dependence of $\epsilon_{\rm el}^{-1}$ using the plasmon-pole model~\cite{Hybersten1986GW,Liang2015Plasmon}.
The plasmon-pole dielectric function reads
\begin{equation} \label{eq:supp_plasmon_pole_model}
    \epsilon_{\rm el}^{-1}(\qw) - 1
    = U_\mb{q} \delta\chi(\qw)
    = \frac{\Omega_\qz^2}{(\omega + i\gamma_\qz)^2 - \omega_\qz^2},
\end{equation}
where $\Omega_\qz$, $\omega_\qz$, and $\gamma_\qz$ are the plasmon strength, frequency, and linewidth, respectively.
These parameters are determined by fitting \Eq{eq:supp_plasmon_pole_model} to the dielectric function calculated at a few frequencies using Eqs.~(\ref{eq:plph_chi0}, \ref{eq:epsilon}).
Compared to the plasmon-pole models used in $GW$ calculations~\cite{Hybersten1986GW}, our model has an additional parameter $\gamma_\qz$ that describes the plasmon lifetime in the electron-hole continuum.
We use the index 0 to refer to plasmons.
Later, it will be accompanied by the indices 1 to $\Nph=3N_{\rm atom}$ for the phonons, where $N_{\rm atom}$ is the number of atoms per unit cell.

\begin{figure}[t]
\centering
\includegraphics[width=0.99\linewidth]{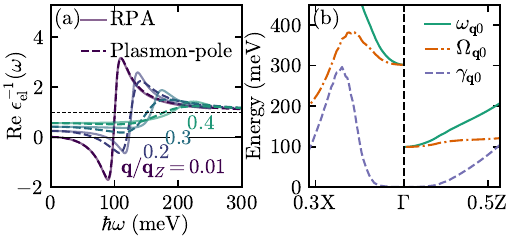}
\caption{
    (a) Free-carrier dielectric functions of $n$-doped anatase TiO$_2$ at $n=1\times10^{19}~\mathrm{cm}^{-3}$ and $T=300~\mathrm{K}$ computed in the RPA using a smearing of 10~meV (solid curves) and fitted with the plasmon-pole model (dashed curves).
    (b) The corresponding plasmon parameters for $q$ points along the X-$\Gamma$-Z line.
}
\label{fig:params}
\end{figure}

To validate the plasmon-pole model in $n$-doped anatase TiO$_2$, we plot the free-carrier dielectric functions computed in the RPA and their plasmon-pole fit in \Fig{fig:params}(a).
For all momenta, the plasmon-pole model gives a good account of the frequency dependence of the dielectric function.
Figure~\ref{fig:params}(b) shows the corresponding plasmon parameters.

\subsection{Phonon Dyson equation}

Now, let us consider the phonons.
The nonadiabatic phonon self-energy due to the doped carriers is~\cite{2010CalandraPhonon,Nomura2015cDFPT,2023BergesPhonon}
\begin{equation} \label{eq:selfen}
    \Pi_{\mu\nu}(\qw)
    = (g_\qmu^{\rm lr})^* \, \delta \chi(\qw) \, g_\qnu^{\rm lr}
\end{equation}
with $\mu, \nu = 1, \cdots, \Nph$ the phonon mode index, and
$g_\qnu^{\rm lr}$ the long-range electron-phonon vertex, which is usually calculated as the sum of the dipolar~\cite{Sjakste2015Frohlich,Verdi2015Frohlich} and quadrupolar~\cite{2019RoyoQuadrupole,2020BruninPiezo1,2020BruninPiezo2,2020JhalaniPiezo,2020ParkPiezo} contributions.
Here, we consider the dominant, dipolar contribution only, which is~\cite{Verdi2015Frohlich}
\begin{equation} \label{eq:eph_lr}
    g_\qnu^{\rm lr}
    = i\frac{4\pi e^2}{Vq} \sum_{\kappa} \frac{1}{\sqrt{M_\kappa}} \frac{\hat{q}\cdot\mb{Z}^*_\kappa \cdot \mb{e}_{\kappa\nu}(\mb{q})}{\epsilon^\infty_{\hat{q}}}.
\end{equation}
where $M_\kappa$ is the mass of the $\kappa$-th atom in the unit cell, $\omega_\qnu$ the frequency of the phonon in the $\nu$-th branch and at wavevector $\mb{q}$, $\mb{Z}^*_\kappa$ the Born effective charge tensor, and $\mb{e}_{\kappa\nu}(\mb{q})$ the phonon eigendisplacement normalized as
\begin{equation}
    \sum_{\kappa\alpha} [e^\alpha_{\kappa\mu}(\mb{q})]^* e^\alpha_{\kappa\nu}(\mb{q}) = \delta_{\mu\nu}.
\end{equation}
(Note that our convention for $g_\qnu^{\rm lr}$ differs from that of Ref.~\cite{Verdi2015Frohlich} by a factor of $1 / \sqrt{2 \omega_\qnu}$.)
The bare phonon Green function is
\begin{equation} \label{eq:D0_phonon}
    D^{0}_{\mu\nu}(\qw) = \delta_{\mu\nu} \frac{1}{(\omega + i \gamma_\qnu)^2 - \omega_\qnu^2}.
\end{equation}
The phonon inverse lifetime $\gamma_\qnu$ can be used to include intrinsic phonon lifetimes.
We note that we consider retarded Green functions, contrary to time-ordered Green functions used, e.g., in Eq.~(134) of Ref.~\cite{2017GiustinoRMP}.

For later use, we also define the long-range electron-plasmon vertex~\cite{Caruso2016Plasmon}
\begin{align} \label{eq:plph_g0_def}
    g^{\rm lr}_\qz
    &= \frac{1}{q} \Biggl[ \frac{1}{2\omega_\qz}\, \frac{V \epsilon^{\infty}_{\hat{q}}}{4\pi e^2}\, \frac{\partial \epsilon_{\rm el}(\qw)}{\partial \omega} \Bigr|_{\omega_\qz} \Biggr]^{-1/2}
    \nnnl
    &= \frac{1}{q} \Biggl[ \frac{1}{2\omega_\qz}\,\frac{1}{q^2\,U_\mb{q}}\, \frac{2\omega_\qz}{\Omega_\qz^2} \Biggr]^{-1/2}
    \nnnl
    &= \Omega_\qz \sqrt{U_\mb{q}}
\end{align}
and the bare plasmon Green function
\begin{equation} \label{eq:D0_plasmon}
    D^{0}_{00}(\qw) = \frac{1}{(\omega + i \gamma_\qz)^2 - \omega_\qz^2}.
\end{equation}
(Again, our convention for the vertex differs from that of Ref.~\cite{Caruso2016Plasmon} by a factor of $1/\sqrt{2\omega_\qz}$.)
By defining the plasmon-phonon coupling amplitude,
\begin{equation} \label{eq:plph_coupling}
    c_\qnu = \frac{g^{\rm lr}_\qz g^{\rm lr}_\qnu}{U_\mb{q}},
\end{equation}
we can rewrite the phonon self-energy~[\Eq{eq:selfen}] as
\begin{align} \label{eq:selfen_rewrite}
    \Pi_{\mu\nu}(\qw)
    &= \frac{1}{U_\mb{q}} (g_\qmu^{\rm lr})^*  \frac{\Omega_\qz^2}{(\omega + i\gamma_\qz)^2 - \omega_\qz^2} g_\qnu^{\rm lr}
    \nnnl
    &= \frac{1}{U_\mb{q}^2} (g_\qmu^{\rm lr} g^{\rm lr}_\qz)^* D^{0}_{00}(\qw) g^{\rm lr}_\qz g_\qnu^{\rm lr}
    \nnnl
    &= c_\qmu^* D^{0}_{00}(\qw) c_\qnu.
\end{align}

The full phonon Green function $D_{\mu\nu}(\qw)$ is the solution of the Dyson equation
\begin{equation} \label{eq:Dyson_phonon}
    D^{-1}_{\mu\nu}(\qw)
    = (D^{0})^{-1}_{\mu\nu}(\qw) - \Pi_{\mu\nu}(\qw).
\end{equation}
The phonon spectral function is
\begin{equation}
    A(\qw) = -\frac{2\omega}{\pi} \sum_{\nu=1}^{\Nph} \Im D_{\nu\nu}(\mb{q},\omega).
\end{equation}
Solving \Eq{eq:Dyson_phonon} requires an $\Nph\times \Nph$ matrix inversion \textit{at each frequency}, and is therefore unsuitable for first-principles calculations involving many phonon modes and wavevectors. Also, it is hard to obtain physical intuition from such a formalism.
In the next section, we discuss how to overcome this limitation and transform this frequency-dependent problem into a frequency-independent problem in the $(\Nph+1)$-dimensional space of a plasmon and $\Nph$ phonons.

\subsection{Plasmon-phonon Green function and dynamical matrix} \label{sec:supp_plph}
We start by defining the plasmon-phonon Green function with indices $\alpha, \alpha' = 0, 1, \cdots, \Nph$:
\begin{equation} \label{eq:D_extended}
    \widetilde{D}^{-1}(\qw)
    = \begin{pmatrix}
        (D^{0}_{00})^{-1} & -c_{\mb{q}1} & -c_{\mb{q}2} & \cdots
        \\
        -c_{\mb{q}1}^* & (D^{0}_{11})^{-1} &  0 & \cdots
        \\
        -c_{\mb{q}2}^* & 0 &  (D^{0}_{22})^{-1} & \cdots
        \\
        \vdots & \vdots & \vdots & \ddots
    \end{pmatrix},
\end{equation}
where we have omitted the $(\qw)$ arguments for the bare Green functions $(D^0_{\alpha\alpha})^{-1}$.
We denote $(\Nph+1)$-dimensional objects with a tilde.
The diagonal entries of $\widetilde{D}^{-1}$ are the inverse of the bare plasmon and phonon Green functions.
The off-diagonal entries are the plasmon-phonon coupling coefficients.

From the Schur complement
$\begin{psmallmatrix}A & B \\ C & D\end{psmallmatrix}^{-1} = \begin{psmallmatrix} \cdots & \cdots \\ \cdots & (D - C A^{-1} B)^{-1} \end{psmallmatrix}$
and by using \Eq{eq:selfen_rewrite},
we find that the lower right block of $\widetilde{D}$ is equal to the interacting phonon Green function:
\begin{equation} \label{eq:Dtilde_munu}
    \widetilde{D}_{\mu\nu}(\qw) = D_{\mu\nu}(\qw) \ \text{for}\ \mu, \nu = 1, \cdots, \Nph.
\end{equation}
Thus, by calculating $\widetilde{D}$ one can recover the full phonon Green function $D$.
The other components of $\widetilde{D}$ are
\begin{subequations} \label{eq:Dtilde_other}
\begin{align}
    \label{eq:Dtilde_00}
    \widetilde{D}_{00}
    &= D^0_{00} + \sum_{\mu,\nu=1}^{\Nph} D^0_{00} c_\qmu D_{\mu\nu} c_\qnu^* D^0_{00},
    \\
    \label{eq:Dtilde_0nu}
    \widetilde{D}_{0\nu}
    &= \sum_{\mu=1}^{\Nph} D^0_{00} c_\qmu D_{\mu\nu},
    \\
    \label{eq:Dtilde_mu0}
    \widetilde{D}_{\mu 0}
    &= \sum_{\nu=1}^{\Nph} D_{\mu\nu} c_{\qnu}^* D^0_{00}.
\end{align}
\end{subequations}

Calculating $\widetilde{D}$ instead of $D$ is conceptually and numerically advantageous due to its simpler frequency dependence.
First, we consider the simple case where the plasmon and phonon lifetimes are infinite, i.e., $\gamma_\qa = 0^+$.
Then, \Eq{eq:D_extended} simplifies to
\begin{equation} \label{eq:D_omega_C}
    \widetilde{D}^{-1}_{\alpha\alpha'}(\qw)
    = (\omega + i0^+)^2 - \widetilde{C}_{\mb{q}\alpha\alpha'},
\end{equation}
where
\begin{equation} \label{eq:D_C_def}
    \widetilde{C}_{\mb{q}}
    = \begin{pmatrix}
        \omega_{\mb{q}0}^2 & c_{\mb{q}1} & c_{\mb{q}2} & \cdots
        \\
        c_{\mb{q}1}^* & \omega_{\mb{q}1}^2 & 0 & \cdots
        \\
        c_{\mb{q}2}^* & 0 & \omega_{\mb{q}2}^2 & \cdots
        \\
        \vdots & \vdots & \vdots & \ddots
    \end{pmatrix}
\end{equation}
is the plasmon-phonon dynamical matrix.

Importantly, $\widetilde{C}_{\mb{q}}$ is frequency independent, and the entire frequency dependence of $\widetilde{D}$ can be reconstructed from the eigenvalues and eigenvectors of $\widetilde{C}_{\mb{q}}$.
Since $\widetilde{C}_{\mb{q}}$ is Hermitian, it yields orthonormal eigenvectors
\begin{equation} \label{eq:D_eigen_Vprime}
    \sum_{\alpha'=0}^{\Nph} \widetilde{C}_{\mb{q} \alpha\alpha'} \widetilde{V}'_{\mb{q} \alpha' \beta}
    = \widetilde{\omega}_{\mb{q}\beta}^2 \widetilde{V}'_{\mb{q} \alpha \beta}.
\end{equation}
Defining
\begin{equation} \label{eq:D_eigen_V}
    \widetilde{V}_{\mb{q}\alpha \beta} = \widetilde{V}'_{\mb{q}\alpha \beta} / \sqrt{2\widetilde{\omega}_\qb}
\end{equation}
and substituting this eigendecomposition into \Eq{eq:D_omega_C}, we get
\begin{align} \label{eq:D_omega_eigen}
    &\widetilde{D}_{\alpha\alpha'}(\qw)
    = \bigl[ (\omega + i0^+)^2 - \widetilde{C}_{\mb{q}} \bigr]_{\alpha\alpha'}^{-1},
    \nnnl
    &= \sum_{\beta=0}^{\Nph} \widetilde{V}_{\mb{q}\alpha \beta}' \widetilde{V}_{\mb{q}\beta \alpha'}^{\prime\dagger} \frac{1}{(\omega + i0^+)^2 - \widetilde{\omega}_\qb^2}
    \nnnl
    &= \sum_{\beta=0}^{\Nph} \widetilde{V}_{\mb{q}\alpha \beta} \widetilde{V}_{\mb{q}\beta \alpha'}^\dagger \frac{2\widetilde{\omega}_\qb}{(\omega + i0^+)^2 - \widetilde{\omega}_\qb^2}.
\end{align}
The full Green function of coupled plasmons and phonons is a linear combination of the Green functions of $\Nph+1$ non-interacting plasmon-phonon hybrid modes with energies $\widetilde{\omega}_\qb$ and eigenvectors $\widetilde{V}_{\mb{q}\alpha \beta}$.

If the plasmons or phonons have a finite lifetime, \Eq{eq:D_omega_C} becomes
\begin{equation} \label{eq:D_omega_C_Gamma}
    \widetilde{D}^{-1}_{\alpha\alpha'}(\qw)
    = \bigl[ (\omega + i \widetilde{\Gamma}_\mb{q})^2 - \widetilde{C}_{\mb{q}} \bigr]_{\alpha\alpha'},
\end{equation}
where $\widetilde{\Gamma}_{\mb{q}\aap} = \delta_{\aap}\, \gamma_\qa$.
In general, $\widetilde{\Gamma}_\mb{q}$ does not commute with $\widetilde{C}_\mb{q}$, and the two matrices cannot be diagonalized simultaneously.
However, the frequency dependence can still be simplified by defining a $2(\Nph+1)\times2(\Nph+1)$ block matrix:
\begin{equation} \label{eq:broad_matrix_extended}
    \bigl[(\omega + i \widetilde{\Gamma}_\mb{q})^2 - \widetilde{C}_\mb{q} \bigr]^{-1}
    = \begin{pmatrix} \one & 0 \end{pmatrix}
    \begin{pmatrix}
        \omega + i \widetilde{\Gamma}_\mb{q} & -\one \\
        -\widetilde{C}_\mb{q} & \omega + i\widetilde{\Gamma}_\mb{q}
    \end{pmatrix}^{-1}
    \begin{pmatrix}0 \\ \one \end{pmatrix}.
\end{equation}
Here, $\one$ is an $(\Nph+1)$-dimensional identity matrix.
This result follows from a block matrix identity
\begin{equation} \label{eq:M_QDQinv}
    \begin{pmatrix}
    A & B \\ C & D\end{pmatrix}^{-1}
    = \begin{pmatrix} \cdots &
    (C-DB^{-1}A)^{-1}
    \\ \cdots & \cdots
    \end{pmatrix}.
\end{equation}

To perform an eigendecomposition of \Eq{eq:broad_matrix_extended}, we first define
\begin{equation}
    e^{i\phi_\qnu} = c_\qnu / \abs{c_\qnu}
\end{equation}
and
\begin{equation} \label{eq:Phi}
    \widetilde{\Phi}_\mb{q} = \mathrm{diag}(1, e^{i\phi_{\mb{q}1}}, \ccdots, e^{i\phi_{\mb{q}\Nph}}),
\end{equation}
which is a unitary transformation that yields a real, symmetric matrix
\begin{equation}
    \widetilde{\Phi}_\mb{q} \widetilde{C}_{\mb{q}} \widetilde{\Phi}_\mb{q}^\dagger
    = \begin{pmatrix}
        \omega_{\mb{q}0}^2 & \abs{c_{\mb{q}1}} & \abs{c_{\mb{q}2}} & \cdots
        \\
        \abs{c_{\mb{q}1}} & \omega_{\mb{q}1}^2 & 0 & \cdots
        \\
        \abs{c_{\mb{q}2}} & 0 & \omega_{\mb{q}2}^2 & \cdots
        \\
        \vdots & \vdots & \vdots & \ddots
    \end{pmatrix}\,.
\end{equation}
Also, since both $\widetilde{\Phi}_\mb{q}$ and $\widetilde{\Gamma}_\mb{q}$ are diagonal, we find
\begin{equation}
    \widetilde{\Phi}_\mb{q} \widetilde{\Gamma}_{\mb{q}} \widetilde{\Phi}_\mb{q}^\dagger
    = \widetilde{\Gamma}_{\mb{q}}.
\end{equation}

Next, we define
\begin{align} \label{eq:broad_eigen_M_def}
    \begin{pmatrix}
        \widetilde{A} & \widetilde{B} \\ -\widetilde{B} & -\widetilde{A}^*
    \end{pmatrix}
    &= \frac{1}{2} \begin{pmatrix} \widetilde{\Phi} & \widetilde{\Phi} \\ \widetilde{\Phi} & -\widetilde{\Phi} \end{pmatrix}
    \begin{pmatrix}
        -i \widetilde{\Gamma} & \one \\
        \widetilde{C} & - i\widetilde{\Gamma}
    \end{pmatrix}
    \begin{pmatrix} \widetilde{\Phi}^\dagger & \widetilde{\Phi}^\dagger \\ \widetilde{\Phi}^\dagger & -\widetilde{\Phi}^\dagger \end{pmatrix}
\end{align}
where
\begin{align}
    \widetilde{A}
    &= \tfrac{1}{2}\bigl( \widetilde{\Phi}\, \widetilde{C}\, \widetilde{\Phi}^\dagger + \one \bigr) -i \widetilde{\Gamma},
    \\
    \widetilde{B}
    &= \tfrac{1}{2}\bigl( \widetilde{\Phi}\, \widetilde{C}\, \widetilde{\Phi}^\dagger - \one \bigr),
\end{align}
are $(\Nph+1)$-dimensional symmetric matrices.
We have omitted the subscript $\mb{q}$ for brevity.
This matrix has the same form as the matrix that appears in the time-dependent Hartree--Fock theory~\cite{McWeenyBook,Bauernschmitt1996TDDFT}.
For a right eigenvector with eigenvalue $\wtl$
\begin{align} \label{eq:supp_broad_AB_1}
    \begin{pmatrix}
        \widetilde{A} & \widetilde{B} \\ -\widetilde{B} & -\widetilde{A}^*
    \end{pmatrix}
    \begin{pmatrix} \wtx \\ \wty \end{pmatrix}
    = \wtl \begin{pmatrix} \wtx \\ \wty \end{pmatrix},
\end{align}
one can find a right eigenvector with eigenvalue $-\wtl^*$
\begin{align}
    \begin{pmatrix}
        \widetilde{A} & \widetilde{B} \\
        -\widetilde{B} & -\widetilde{A}^*
    \end{pmatrix}
    \begin{pmatrix} \wty^* \\ \wtx^* \end{pmatrix}
    = -\wtl^* \begin{pmatrix} \wty^* \\ \wtx^* \end{pmatrix},
\end{align}
as well as left eigenvectors
\begin{align}
    \begin{pmatrix} \wtx^\intercal & -\wty^\intercal \end{pmatrix}
    \begin{pmatrix}
        \widetilde{A} & \widetilde{B} \\ -\widetilde{B} & -\widetilde{A}^*
    \end{pmatrix}
    = \wtl \begin{pmatrix} \wtx^\intercal & -\wty^\intercal \end{pmatrix},
\end{align}
\begin{align}
    \begin{pmatrix} -\wty^{\intercal*} & \wtx^{\intercal*} \end{pmatrix}
    \begin{pmatrix}
        \widetilde{A} & \widetilde{B} \\ -\widetilde{B} & -\widetilde{A}^*
    \end{pmatrix}
    = -\wtl^* \begin{pmatrix} -\wty^{\intercal*} & \wtx^{\intercal*} \end{pmatrix}.
\end{align}
From the biorthogonality of left and right eigenvectors, we may orthonormalize $\wtx$ and $\wty$ to satisfy
\begin{equation}
    \begin{pmatrix}
        \wtx^\intercal & -\wty^\intercal \\ -\wty^{\intercal*} & \wtx^{\intercal*}
    \end{pmatrix}
    \begin{pmatrix}
        \wtx & \wty^* \\ \wty & \wtx^*
    \end{pmatrix}
    = \begin{pmatrix}
        1 & 0 \\ 0 & 1
    \end{pmatrix}.
\end{equation}
Hence, eigendecomposition yields
\begin{equation} \label{eq:broad_M_eigen}
    \begin{pmatrix}
        \wtA & \wtB \\ -\wtB & -\wtA^*
    \end{pmatrix}
    = \begin{pmatrix}
        \wtX & \wtY^* \\ \wtY & \wtX^*
    \end{pmatrix}
    \begin{pmatrix}
        \wtL & 0 \\ 0 & -\wtL^*
    \end{pmatrix}
    \begin{pmatrix}
        \wtX^\intercal & -\wtY^\intercal \\ -\wtY^{\intercal*} & \wtX^{\intercal*}
    \end{pmatrix},
\end{equation}
where $\wtL_{\beta\beta'} = \wtl_\beta \delta_{\beta\beta'}$ is a diagonal matrix, and $\wtX$ and $\wtY$ are square matrices, whose $\beta^{\rm th}$ columns are $\wtx$ and $\wty$ in \Eq{eq:supp_broad_AB_1} for $\wtl = \wtl_\beta$.
We choose the eigenvalues $\wtl_\beta$ and $-\wtl_\beta^*$ such that $\mathrm{Re}\wtl_\beta > 0$.
We also used the biorthogonality of the left and right eigenvectors related to different $(\wtl_\beta,\, -\wtl_\beta^*)$ pairs.

By substituting \Eqs{eq:broad_eigen_M_def} and \eqref{eq:broad_M_eigen}
into \Eq{eq:broad_matrix_extended}, we find
\begin{align} \label{eq:D_using_Z_broad}
    &\bigl[(\omega + i \widetilde{\Gamma})^2 - \widetilde{C} \bigr]^{-1}
    \nnnl
    &= \frac{1}{2} \begin{pmatrix} \wtP^\dagger & \wtP^\dagger \end{pmatrix}
    \begin{pmatrix}
        \wtX & \wtY^* \\ \wtY & \wtX^*
    \end{pmatrix}
    \begin{pmatrix}
        \omega - \wtL & 0 \\ 0 & \omega +\wtL^*
    \end{pmatrix}^{-1}
    \nnnl
    &\quad \times
    \begin{pmatrix}
        \wtX^\intercal & -\wtY^\intercal \\ -\wtY^{\intercal*} & \wtX^{\intercal*}
    \end{pmatrix}
    \begin{pmatrix} \wtP \\ -\wtP \end{pmatrix}
    \nnnl
    &= \wtP^\dagger \begin{pmatrix}
        \wtZ & \wtZ^*
    \end{pmatrix}
    \begin{pmatrix}
        \frac{1}{\omega - \wtL} & 0 \\ 0 & \frac{1}{\omega +\wtL^*}
    \end{pmatrix}
    \begin{pmatrix} \wtZ^\intercal \\ -\wtZ^{\intercal*} \end{pmatrix} \wtP,
\end{align}
where we defined
\begin{equation}
    \wtZ = \frac{1}{\sqrt{2}} \bigl( \wtX + \wtY \bigr).
\end{equation}

Finally, we define
\begin{equation}
    \wtV^{\rm L}_{\alpha\beta}
    = e^{-i \wtp_\alpha} \wtZ_{\alpha\beta},\quad
    \wtV^{\rm R}_{\alpha\beta}
    = e^{-i \wtp_\alpha} \wtZ^*_{\alpha\beta},
\end{equation}
for which we set $\wtp_0=0$ [Eq.~\eqref{eq:Phi}], and write the real and imaginary parts of $\wtl_\beta$ as
\begin{equation}
    \wtl_\beta = \wtw_\beta -i\wtg_\beta.
\end{equation}
Here, $\wtw_\beta$ and $\wtg_\beta$ are the energy and linewidth of the hybrid mode.
By reinstating the $\mb{q}$ subscripts, we obtain our final result for the plasmon-phonon Green function:
\begin{align} \label{eq:D_eigen_broad}
    \wtD_{\alpha\alpha'}(\qw)
    &= \sum_{\beta=0}^{\Nph}
    \Biggl[
    \frac{e^{-i\phi_\qa} \wtZ_{\mb{q}\alpha\beta} \wtZ_{\mb{q}\alpha'\beta} e^{i\phi_\qap}}{\omega - \wtw_\qb + i \wtg_\qb}
    \nnnl
    &\quad+ \Biggl( \frac{e^{i\phi_\qa} \wtZ_{\mb{q}\alpha\beta} \wtZ_{\mb{q}\alpha'\beta} e^{-i\phi_\qap}}{-\omega - \wtw_\qb + i \wtg_\qb} \Biggr)^*
    \Biggr]
    \nnnl
    &= \sum_{\beta=0}^{\Nph}
    \frac{\wtV^{\rm L}_{\mb{q}\alpha\beta} \wtV^{\rm R\dagger}_{\mb{q}\beta\alpha'}}{\omega - \wtw_\qb + i \wtg_\qb}
    + (\omega \to -\omega,\ \alpha \leftrightarrow \alpha')^*\,.
\end{align}

In the limit $\gamma_\qa \to 0^+$, we use \Eqs{eq:D_eigen_Vprime} and \eqref{eq:D_eigen_V} to find
\begin{align}
    &\bigl[(\omega + i 0^+)^2 - \widetilde{C} \bigr]^{-1}
    \nnnl
    &= \wtV' \frac{1}{\sqrt{2\widetilde{\Omega}}} \Bigl[\bigl( \omega + i 0^+ -  \widetilde{\Omega}  \bigr)^{-1}
    - \bigl(\omega + i 0^+ + \widetilde{\Omega} \bigr)^{-1} \Bigr] \frac{1}{\sqrt{2\widetilde{\Omega}}} \wtV^{\prime\dagger}
    \nnnl
    &= \wtV \Bigl[\bigl( \omega + i 0^+ -  \widetilde{\Omega}  \bigr)^{-1}
    - \bigl(\omega + i 0^+ + \widetilde{\Omega} \bigr)^{-1} \Bigr] \wtV^{\dagger}
    \nnnl
    &= \begin{pmatrix}
        \wtV & \wtV
    \end{pmatrix}
    \begin{pmatrix}
        \frac{1}{\omega + i0^+ - \widetilde{\Omega}} & 0 \\ 0 & \frac{1}{\omega + i0^+ + \widetilde{\Omega}}
    \end{pmatrix}
    \begin{pmatrix} \wtV^\dagger \\ -\wtV^\dagger \end{pmatrix},
\end{align}
where $\widetilde{\Omega}_{\beta\beta'} = \wtw_\beta \delta_{\beta\beta'}$.
Comparing this equation with \Eq{eq:D_using_Z_broad}, we find
\begin{equation} \label{eq:broad_to_nobroad}
    \wtV^{\rm L} = \wtV^{\rm R} = \wtV,\quad \wtg_\qb = 0^+.
\end{equation}
With these equations, \Eq{eq:D_eigen_broad} reduces to \Eq{eq:D_omega_eigen}.

\subsection{Static approximation}
For calculations in the static approximation, we fixed the plasmon frequency to a very large value ($\omega_\qz=\omega^*=5~\mathrm{eV}$; the result is insensitive to $\omega^*$ as long as $\omega \ll \omega^*$), fixed the broadening to zero ($\gamma_\qz=0$), and determined the plasmon strength $\Omega_\qz$ using the dielectric function at zero frequency:
\begin{equation}
    \Omega_\qz = \omega^* \sqrt{1 - \epsilon_{\rm el}^{-1}(\mb{q},0) }.
\end{equation}
This choice of the plasmon-pole model parameters yields a frequency independent dielectric function in the low energy region ($\omega \ll \omega^*$):
\begin{equation}
    1 + \frac{\Omega_\qz^2}{\omega^2 - (\omega^*)^2}
    \approx \epsilon_{\rm el}^{-1}(\mb{q},0).
\end{equation}
We used these parameters to construct the plasmon-phonon dynamical matrix and diagonalized it to obtain the statically-screened phonons.
This scheme generates the phonon dispersion in the static approximation $\Pi(\qw) \approx \Pi(\mb{q},0)$.

\subsection{Generalization to the mutipole model}
If the frequency-dependent susceptibility has complex frequency dependence that the plasmon-pole approximation cannot capture, one can still use our method, generalized by adopting the multipole method~\cite{Leon2021Multipole}.
In the multipole method, the dielectric function is modeled as a sum over multiple poles:
\begin{equation} \label{eq:supp_multipole}
    \epsilon_{\rm el}^{-1}(\qw) - 1
    = \sum_{p=1}^{\Npl} \frac{\bar{\Omega}_\qp^2}{(\omega + i\bar{\gamma}_\qp)^2 - \bar{\omega}_\qp^2}.
\end{equation}
We denote the plasmon parameters with a bar over the variable.
Correspondingly, we define an $\Npl$-dimensional bare plasmon Green function
\begin{equation} \label{eq:D0_plasmon_multipole}
    \bar{D}^{0}_{pp'}(\qw) = \frac{1}{(\omega + i \bar{\gamma}_\qp)^2 - \bar{\omega}_\qp^2} \delta_{pp'}.
\end{equation}
The phonon self-energy, which was given by \Eq{eq:selfen_rewrite} in the case of a single pole, now reads
\begin{align} \label{eq:selfen_rewrite_multipole}
    \Pi_{\mu\nu}(\qw)
    = \sum_{p,p'=1}^{\Npl} c_\qmu^* D^{0}_{pp'}(\qw) c_\qnu.
\end{align}
The plasmon-phonon Green function is now an $(\Nph + \Npl) \times (\Nph + \Npl)$ matrix:
\begin{multline} \label{eq:D_extended_multipole}
    \widetilde{D}^{-1}(\qw)
    \\
    \begingroup
    \setlength\arraycolsep{-0.5pt}
    = \begin{pmatrix}
        (\bar{D}^{0}_{11})^{-1} & & 0 & -c_{\mb{q}1} & \ccdots & -c_{\mb{q}\Nph}
        \\
         & \ddots &  & \vdots & \vdots & \ccdots
        \\
        0 & & (\bar{D}^{0}_{\Npl\Npl})^{-1} & -c_{\mb{q}1} & \ccdots & -c_{\mb{q}\Nph}
        \\
        -c_{\mb{q}1}^* & \ccdots & -c_{\mb{q}1}^* & (D^{0}_{11})^{-1} &  0 & \ccdots
        \\
        \vdots & \vdots & \vdots & \vdots & \vdots & \ddots
        \\
        -c_{\mb{q}\Nph}^* & \ccdots & -c_{\mb{q}\Nph}^* & 0 &  & (D^{0}_{\Nph\Nph})^{-1} 
    \end{pmatrix}.
    \endgroup
\end{multline}
All of the results derived in \Sec{sec:supp_plph} can be straightforwardly generalized by using
\begin{equation} \label{eq:D_C_def_multipole}
    \widetilde{C}_{\mb{q}}
    = \begin{pmatrix}
        \bar{\omega}_{\mb{q}1}^2 &  & 0 & c_{\mb{q}1} & \ccdots & c_{\mb{q}\Nph}
        \\
         & \ddots & & \vdots  & \vdots & \vdots
        \\
        0 &  & \bar{\omega}_{\mb{q}\Npl}^2 & c_{\mb{q}1} & \ccdots & c_{\mb{q}\Nph}
        \\
        c_{\mb{q}1}^* & \ccdots & c_{\mb{q}1}^* & \omega_{\mb{q}1}^2 &  & 0
        \\
        \vdots & \vdots & \vdots &  & \ddots & 
        \\
        c_{\mb{q}\Nph}^* & \ccdots & c_{\mb{q}\Nph}^* & 0 & &  \omega_{\mb{q}\Nph}^2
    \end{pmatrix}
\end{equation}
and
\begin{equation} \label{eq:D_Gamma_def_multipole}
    \widetilde{\Gamma}_{\mb{q}}
    = \begin{pmatrix}
        \bar{\gamma}_{\mb{q}1} &  & 0 & 0 & \cdots & 0
        \\
        & \ddots & & \vdots & \ddots & \vdots
        \\
        0 &  & \bar{\gamma}_{\mb{q}\Npl} & 0 & \cdots & 0
        \\
        0 & \cdots & 0 & \gamma_{\mb{q}1} & & 0
        \\
        \vdots & \ddots & \vdots & & \ddots & 
        \\
        0 & \cdots & 0 & 0 & & \gamma_{\mb{q}\Nph}
    \end{pmatrix}.
\end{equation}
The plasmon-phonon hybrids can be obtained by diagonalizing an $(\Npl+\Nph)\times(\Npl+\Nph)$ matrix for the case of infinitesimal linewidths, and a $2(\Npl+\Nph)\times 2(\Npl+\Nph)$ matrix for the case of finite linewidths.

\section{Electron-plasmon-phonon coupling and dielectric function} \label{sec:supp_eph}

The description in terms of plasmon-phonon hybrid modes yields a simple sum-over-modes expression for many quantities.
As we have shown for the phonon Green function in the previous section, here we derive analogous expressions for the electron-plasmon-phonon coupling and the dielectric function.

\subsection{Electron-plasmon-phonon coupling} \label{sec:supp_eph_deriv}
We begin by defining the electron-plasmon and electron-phonon vertices.
Since we have neglected local field effects, the electrons couple to the plasmons only via the long-range potential.
The electron-plasmon vertex~\cite{Caruso2016Plasmon} is
\begin{equation} \label{eq:epp_g_pl}
    g_{mn0}(\mb{k},\mb{q}) = g^{\rm lr}_\qz \braket{u_\mkq}{u_\nk}.
\end{equation}
The unscreened electron-phonon vertex is the sum of the short- and long-range contributions:
\begin{equation} \label{eq:epp_g_ph}
    g_{mn\nu}(\mb{k},\mb{q}) = g^{\rm sr}_{mn\nu}(\mb{k},\mb{q})
    + g_\qnu^{\rm lr} \braket{u_\mkq}{u_\nk}.
\end{equation}
By ``unscreened'', we mean specifically that the free-carrier screening is not included; the dielectric screening of the undoped system is taken into account in the static density-functional perturbation theory (DFPT).
This electron-phonon vertex is dynamically screened by the plasmons as
\begin{align} \label{eq:plph_gnu_scr}
    &g_{mn\nu}^{\rm scr}(\mb{k},\qw)
    \nnnl
    &= g^{\rm sr}_{mn\nu}(\mb{k},\mb{q})
    + \epsilon_{\rm el}^{-1}(\qw) g_\qnu^{\rm lr} \braket{u_\mkq}{u_\nk}
    \nnnl
    &= g_{mn\nu}(\mb{k},\mb{q}) +  \frac{\Omega_\qz^2}{(\omega + i\gamma_\qz)^2 - \omega_\qz^2} g_\qnu^{\rm lr} \braket{u_\mkq}{u_\nk}
    \nnnl
    &= g_{mn\nu}(\mb{k},\mb{q}) + \frac{1}{U_\mb{q}} g_{mn0}(\mb{k},\mb{q}) D^0_{00}(\omega)
    g_\qz^{\rm lr} g_\qnu^{\rm lr}
    \nnnl
    &= g_{mn\nu}(\mb{k},\mb{q}) + g_{mn0}(\mb{k},\mb{q}) D^0_{00}(\omega) c_\qnu.
\end{align}
The first term in the second line is the unscreened short-range part, and the second term accounts for the screening of the long-range part.
Since plasmons are long-range charge fluctuations, they only screen the latter.
In the second equality, we used our plasmon-pole model [\Eq{eq:supp_plasmon_pole_model}].
In the third equality, we used \Eqs{eq:plph_g0_def}, \eqref{eq:D0_plasmon}, and \eqref{eq:epp_g_pl}.
To get to the last line, we used \Eq{eq:plph_coupling}.

The effective electron-electron interaction between the states $(n,\mb{k})$ and $(m,\mb{k}+\mb{q})$ is the sum of the electron and phonon contributions, which are~\cite{2017GiustinoRMP,Filip2021}
\begin{equation} \label{eq:eph_W_plph}
    W^\plph_{mn\mb{k}}(\qw)
    = W^{\rm el}_{mn\mb{k}}(\qw)
    + W^{\rm ph}_{mn\mb{k}}(\qw),
\end{equation}
\begin{align} \label{eq:eph_W_el}
    W^{\rm el}_{mn\mb{k}}(\qw)
    &= U_\mb{q} \epsilon_{\rm el}^{-1}(\qw) \abs{\braket{u_\mkq}{u_\nk}}^2,
    \\
    &= U_\mb{q}
    + g_{mn0}(\mb{k},\mb{q}) D^0_{00}(\qw) [g_{mn0}(\mb{k},\mb{q})]^*, \nonumber
\end{align}
\begin{equation} \label{eq:eph_W_ph}
    W^{\rm ph}_{mn\mb{k}}(\qw)
    = \sum_{\mu\nu} g_{mn\mu}^{\rm scr}(\mb{k},\mb{q},\omega) D_{\mu\nu}(\qw) [g_{mn\nu}^{\rm scr}(\mb{k},\mb{q},\omega)]^*.
\end{equation}

Now, we show that rewriting \Eq{eq:eph_W_plph} in terms of the plasmon-phonon Green function [\Eq{eq:D_extended}] yields
\begin{equation} \label{eq:eph_W_in_Dtilde}
    W^\plph_{mn\mb{k}}(\qw)
    = U_\mb{q} + \sum_{\alpha,\alpha' = 0}^{\Nph} g_{mn\alpha} \widetilde{D}_{\aap}(\qw) g_{mn\alpha'}^*.
\end{equation}
This equation means that the electron-plasmon-phonon interaction is described by the plasmon-phonon Green function with the \textit{unscreened} electron-plasmon and electron-phonon vertices.
To derive \Eq{eq:eph_W_in_Dtilde}, we compute the sum over $\alpha$ and $\alpha'$ on the right-hand side by dividing it into four cases, depending on whether the indices are zero or not.
First, using \Eq{eq:Dtilde_00}, the $\alpha=\alpha'=0$ contribution becomes
\begin{align} \label{eq:eph_deriv_1}
    &g_{mn0} \widetilde{D}_{00}(\qw) g_{mn0}^*
    \\
    &= g_{mn0} \Bigl[ D^0_{00} + \sum_{\mu,\nu=1}^{\Nph} D^0_{00} c_\qmu D_{\mu\nu} c_\qnu^* D^0_{00} \Bigr]g_{mn0}^*
    \nnnl
    &= W^{\rm el}_{mn\mb{k}}(\qw) - U_\mb{q} \nnnl
    &\ + \sum_{\mu,\nu=1}^{\Nph} [g_{mn\mu}^{\rm scr} - g_{mn\mu}] D_{\mu\nu} [g_{mn\nu}^{\rm scr} - g_{mn\nu}]^*\,,
    \nonumber
\end{align}
where we have used Eq.~\eqref{eq:plph_gnu_scr} in the second equality.
With \Eq{eq:Dtilde_0nu}, the $\alpha = 0$, $\alpha' = \nu \neq 0$ term becomes
\begin{align} \label{eq:eph_deriv_2}
    \sum_{\nu=1}^\Nph g_{mn0} \widetilde{D}_{0\nu}(\qw) g_{mn\nu}^*
    &= \sum_{\nu=1}^\Nph g_{mn0} \sum_{\mu=1}^{\Nph} D^0_{00} c_\qmu D_{\mu\nu} g_{mn\nu}^*
    \nnnl
    &= \sum_{\mu,\nu=1}^{\Nph} [g_{mn\mu}^{\rm scr} - g_{mn\mu}] D_{\mu\nu} g_{mn\nu}^*.
\end{align}
Similarly, using \Eq{eq:Dtilde_mu0}, the $\alpha = \mu \neq 0$, $\alpha' = 0$ term becomes
\begin{align} \label{eq:eph_deriv_3}
    \sum_{\mu=1}^\Nph g_{mn\mu} \widetilde{D}_{\mu0}(\qw) g_{mn0}^*
    &= \sum_{\mu=1}^\Nph g_{mn\mu} \sum_{\nu=1}^{\Nph} D_{\mu\nu} c_{\qnu}^* D^0_{00} g_{mn0}^*
    \nnnl
    &= \sum_{\mu,\nu=1}^{\Nph} g_{mn\mu} D_{\mu\nu} [g_{mn\nu}^{\rm scr} - g_{mn\nu}]^*.
\end{align}
Finally, from \Eq{eq:Dtilde_munu}, the $\alpha = \mu \neq 0$, $\alpha' = \nu \neq 0$ contribution becomes
\begin{equation} \label{eq:eph_deriv_4}
    \sum_{\mu,\nu=1}^{\Nph} g_{mn\mu} \widetilde{D}_{\mu\nu}(\qw) g_{mn\nu}^*
    = \sum_{\mu,\nu=1}^{\Nph} g_{mn\mu} D_{\mu\nu} g_{mn\nu}^*.
\end{equation}
Adding Eqs.~(\ref{eq:eph_deriv_1}-\ref{eq:eph_deriv_4}) gives \Eq{eq:eph_W_in_Dtilde}.

Finally, we rewrite \Eq{eq:eph_W_in_Dtilde} using the eigenbasis of the plasmon-phonon hybrids.
If $\gamma_\qa \neq 0$, we use \Eq{eq:D_eigen_broad} to find
\begin{multline} \label{eq:W_plph_broad}
    W^\plph_{mn\mb{k}}(\qw)
    = U_\mb{q} \\
    + \sum_{\beta=0}^{\Nph} \frac{\widetilde{g}^{\rm L}_{mn\beta}(\mb{k},\mb{q}) [\widetilde{g}^{\rm R}_{mn\beta}(\mb{k},\mb{q})]^*}{\omega - \wtw_\qb + i \wtg_\qb}
    + (\omega \to -\omega)^*,
\end{multline}
where we have left and right coupling vertices between the electrons and plasmon-phonon hybrids
\begin{align} \label{eq:eph_gtilde_broad}
    \widetilde{g}^{\rm L/R}_{mn\beta}(\mb{k},\mb{q})
    &= \sum_{\alpha=0}^\Nph g_{mn\alpha}(\mb{k},\mb{q}) \widetilde{V}^{\rm L/R}_{\mb{q}\alpha \beta}.
\end{align}

The formula in the zero-broadening limit follows by applying the conditions in \Eq{eq:broad_to_nobroad}, which leads to
\begin{align} \label{eq:eph_gtilde_nobroad}
    \widetilde{g}^{\rm L/R}_{mn\beta}(\mb{k},\mb{q})
    = \widetilde{g}_{mn\beta}(\mb{k},\mb{q})
    &= \sum_\alpha g_{mn\alpha}(\mb{k},\mb{q}) \widetilde{V}_{\mb{q}\alpha\beta},
\end{align}
\begin{equation} \label{eq:W_plph_nobroad}
    W^\plph_{mn\mb{k}}(\qw)
    = U_\mb{q} + \sum_{\beta=0}^\Nph \abs{\widetilde{g}_{mn\beta}(\mb{k},\mb{q})}^2 \frac{2\widetilde{\omega}_\qb}{(\omega + i0^+)^2 - \widetilde{\omega}_\qb^2}.
\end{equation}

Equations \eqref{eq:W_plph_broad} and \eqref{eq:W_plph_nobroad} directly generalize the electron-electron interaction mediated by phonons computed from static DFPT, which has the exact same structure~\cite{2017GiustinoRMP,Filip2021}:
\begin{equation} \label{eq:W_ph}
    W^{\rm ph}_{mn\mb{k}}(\qw)
    = \sum_{\nu=1}^\Nph \abs{g_{mn\nu}(\mb{k},\mb{q})}^2 \frac{2\omega_\qnu}{(\omega + i0^+)^2 - \omega_\qnu^2}.
\end{equation}
Thanks to this analogous form, one can easily generalize formulas and computational workflows for electron-phonon calculations to deal with electron-plasmon-phonon coupling.

An example is the electron self-energy due to the scattering with the plasmon-phonon hybrids.
The Fan--Migdal electron-phonon self-energy is~\cite{MahanBook,2017GiustinoRMP}
\begin{align} \label{eq:el_sigma_ph}
    &\mathrm{Im} \Sigma^{\rm ph}_\nk(\omega)
    \\
    &= \mathrm{Im} \frac{1}{N_q} \sum_{m\mb{q}} \sum_{\nu=1}^{\Nph} \sum_{\pm}
    \frac{\abs{g_{mn\nu}(\mb{k},\mb{q})}^2 \bigl( f^\pm_\mkq + n_\qnu \bigr)}{\omega - \veps_\mkq \pm \omega_\qnu + i 0^+},
    \nonumber
\end{align}
where $N_q$ is the number of sampled $q$ points, $f^+_\mkq$ the Fermi--Dirac occupation at energy $\veps_\mkq$, $f^-_\mkq = 1 - f^+_\mkq$, and $n_\qnu = n(\omega_\qnu) = 1 / \bigl(e^{\omega_\qnu /k_{\rm B} T} - 1\bigr)$ the Bose--Einstein occupation of the phonon mode.

The Fan--Migdal self-energy coming from the plasmon-phonon hybrids with zero broadening has the exact form
\begin{align} \label{eq:el_sigma_plph_nobroad}
    &\mathrm{Im}\Sigma^{\plph}_\nk(\omega)
    \\
    &= \mathrm{Im} \frac{1}{N_q} \sum_{m\mb{q}} \sum_{\beta=0}^{\Nph} \sum_{\pm}
    \frac{\abs{\widetilde{g}_{mn\beta}(\mb{k},\mb{q})}^2 \bigl( f^\pm_\mkq + \widetilde{n}_\qb \bigr)}{\omega - \veps_\mkq \pm \widetilde{\omega}_\qb + i 0^+}\,,
    \nonumber
\end{align}
where the only difference is that all the phonon-related quantities are replaced with the corresponding plasmon-phonon quantities.
For example, the relevant bosonic occupation factor is that of the plasmon-phonon hybrid: $\widetilde{n}_\qb = 1 / \bigl(e^{\widetilde{\omega}_\qb /k_{\rm B} T} - 1\bigr)$.
For the more general finite-broadening case, we find
\begin{align} \label{eq:el_sigma_plph}
    &\mathrm{Im} \Sigma^{\plph}_\nk(\omega)
    = \mathrm{Im}\frac{1}{N_q} \sum_{m\mb{q}} \bigl[ f_\mkq + n(\veps_\mkq - \omega) \bigr] \nnnl
    &\times \sum_{\beta=0}^{\Nph} 
    \Biggl(\frac{[\widetilde{g}^{\rm L}_{mn\beta}(\mb{k},\mb{q})]^* \widetilde{g}^{\rm R}_{mn\beta}(\mb{k},\mb{q})}{\omega - \veps_\mkq + \wtw_\qb + i \wtg_\qb} 
    \nnnl
    &\qquad\qquad -
    \frac{\widetilde{g}^{\rm L}_{mn\beta}(\mb{k},\mb{q}) [\widetilde{g}^{\rm R}_{mn\beta}(\mb{k},\mb{q})]^*}{\omega - \veps_\mkq - \wtw_\qb + i \wtg_\qb} \Biggr).
\end{align}
(See \Sec{sec:supp_sigma_der} for the derivation.)
Equation~\eqref{eq:el_sigma_plph_nobroad} at the on-shell energy $\omega = \veps_\nk$ is Eq.~\eqref{eq:el_sigma} of the main text.
Equations \eqref{eq:el_sigma_plph_nobroad} and \eqref{eq:el_sigma_plph} do not involve any frequency integration and are therefore well suited for \textit{ab initio} calculations using a dense $q$-point grid.

\subsection{Dielectric function} \label{sec:supp_dielec}
The total dielectric function of electrons and phonons can be calculated from the long-range part of the effective electron-electron interaction~\cite{MahanBook}.
We write the long-range electron-electron interaction as $W_{\mb{0}\mb{0}}(\qw)$, as it corresponds to the $\mb{G}=\mb{G'}=\mb{0}$ component of the electron-electron interaction matrix~\cite{Hybersten1986GW}.
Analogously to Eqs.~(\ref{eq:eph_W_plph}-\ref{eq:eph_W_ph}), the long-range component of the electron-electron interaction are defined in terms of the long-range electron-plasmon and electron-phonon vertices~\cite{MahanBook}:
\begin{equation} \label{eq:eph_W00_plph}
    \frac{4\pi e^2}{q^2 V} \epsilon^{-1}_{\plph}(\qw)
    = W^\plph_{\mb{00}}(\qw)
    = W^{\rm el}_{\mb{00}}(\qw)
    + W^{\rm ph}_{\mb{00}}(\qw),
\end{equation}
\begin{align} \label{eq:eph_W00_el}
    W^{\rm el}_{\mb{00}}(\qw)
    &= U_\mb{q}
    + g^{\rm lr}_\qz D^0_{00}(\qw) (g^{\rm lr}_\qz)^*,
\end{align}
\begin{equation} \label{eq:eph_W00_ph}
    W^{\rm ph}_{\mb{00}}(\qw)
    = \sum_{\mu\nu} g_\qmu^{\rm lr,\,scr}(\omega) D_{\mu\nu}(\qw) [g_\qnu^{\rm lr,\, scr}(\omega)]^*,
\end{equation}
where $g_\qnu^{\rm lr,\, scr}(\omega)$ is defined via \Eq{eq:plph_gnu_scr} with $g_{mn\nu}(\mb{k},\mb{q})$ and $g_{mn0}(\mb{k},\mb{q})$ replaced with $g^{\rm lr}_\qnu$ and $g^{\rm lr}_\qz$, respectively.
Following the procedure of \Sec{sec:supp_eph_deriv} that led to \Eq{eq:eph_W_in_Dtilde}, one can show that $W^\plph_{\mb{00}}(\qw)$ can be written in terms of the plasmon-phonon Green function:
\begin{equation}
    W^\plph_{\mb{00}}(\qw)
    = U_\mb{q} + \sum_{\alpha,\alpha' = 0}^{\Nph} g^{\rm lr}_\qa \widetilde{D}_{\aap}(\qw) (g^{\rm lr}_\qap)^*.
\end{equation}

For the case of finite plasmon broadening~[\Eqs{eq:W_plph_broad} and \eqref{eq:eph_gtilde_broad}], we define
\begin{align} \label{eq:dielec_g_lr}
    \widetilde{g}^{\rm lr,\,L/R}_{\qb}
    &= \sum_{\alpha=0}^\Nph g_{\qa}^{\rm lr} \widetilde{V}^{\rm L/R}_{\mb{q}\alpha \beta}
\end{align}
and find
\begin{align} \label{eq:dielec_W_lr}
    W^\plph_{\mb{0}\mb{0}}(\qw)
    = U_\mb{q} + \sum_{\beta=0}^{\Nph} \frac{\widetilde{g}^{\rm lr,\,L}_{\qb} (\widetilde{g}^{\rm lr,\,R}_{\qb})^*}{\omega - \wtw_\qb + i\wtg_\qb}
    + (\omega \to -\omega)^*.
\end{align}
The corresponding inverse dielectric function is
\begin{align} \label{eq:dielec_epsil_broad}
    &\epsilon_{\plph}^{-1}(\qw)
    = \frac{q^2 V}{4\pi e^2} W^{\plph}_{\mb{0}\mb{0}}(\qw)
    \nnnl
    &= \frac{1}{\epsilon_{\hat{q}}^\infty} \left[1
    + \frac{1}{U_\mb{q}}
    \sum_{\beta=0}^{\Nph} \frac{\widetilde{g}^{\rm lr,\,L}_{\qb} (\widetilde{g}^{\rm lr,\,R}_{\qb})^*}{\omega - \wtw_\qb + i\wtg_\qb}
    + (\omega \to -\omega)^* \right].
\end{align}

In the case of infinitesimal plasmon linewidths, we analogously define
\begin{equation}
    \widetilde{g}^{\rm lr}_{\qb}
    = \sum_\alpha g^{\rm lr}_\qa
    \widetilde{V}_{\mb{q}\alpha\beta}
\end{equation}
and find
\begin{equation}
    W^\plph_{\mb{0}\mb{0}}(\qw)
    = U_\mb{q} + \sum_{\beta=0}^\Nph \frac{2\widetilde{\omega}_\qb \bigl| \widetilde{g}_{\qb}^{\rm lr} \bigr|^2}{(\omega + i0^+)^2 - \widetilde{\omega}_\qb^2},
\end{equation}
\begin{align} \label{eq:dielec_epsil_nobroad}
    &\epsilon_{\plph}^{-1}(\qw)
    = \frac{q^2 V}{4\pi e^2} W^{\plph}_{\mb{0}\mb{0}}(\qw)
    \nnnl
    &= \frac{1}{\epsilon_{\hat{q}}^\infty} \left[
    1+ \frac{1}{U_\mb{q}}
    \sum_{\beta=0}^\Nph \frac{2\widetilde{\omega}_\qb \bigl| \widetilde{g}_{\qb}^{\rm lr} \bigr|^2}{(\omega + i0^+)^2 - \widetilde{\omega}_\qb^2}\right].
\end{align}

\begin{figure*}[tb]
\centering
\includegraphics[width=0.99\linewidth]{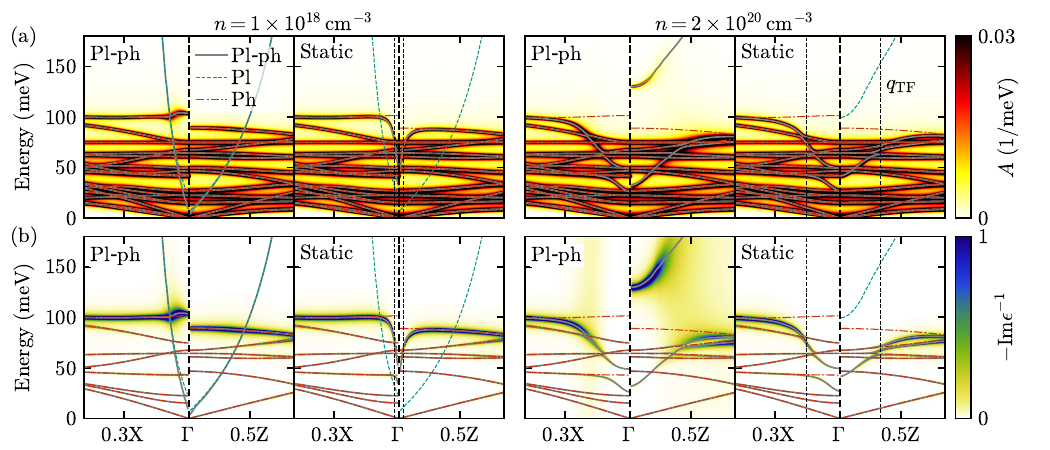}
\caption{
    (a) Phonon spectral function and (b) inverse dielectric function of $n$-doped anatase TiO$_2$.
    The second and fourth columns show the results obtained with the static approximation, while the first and third columns show the results of nonadiabatic plasmon-phonon calculations.
    The curves and lines denote the same quantities as in \Fig{fig:spectral}.
}
\label{fig:spectral_other_doping}
\end{figure*}

Figure~\ref{fig:spectral_other_doping} shows the comparison between our plasmon-phonon method and the static approximation for the phonon spectral function (\Fig{fig:spectral_other_doping}(a)) and the electron loss function (\Fig{fig:spectral_other_doping}(b)) for the low-doping and high-doping cases.
The result for the intermediate-doping case is shown in \Fig{fig:spectral}.

\subsection{Derivation of the electron self-energy} \label{sec:supp_sigma_der}

Here, we derive \Eqs{eq:el_sigma_plph_nobroad} and \eqref{eq:el_sigma_plph}, the electron self-energy induced by the plasmon-phonon hybrids.
To deal with finite-temperature effects, we use the Keldysh formalism~\cite{StefanucciBook}.
All frequency-dependent objects considered so far were the retarded component of the Green functions; here, we denote them by superscript R.
In the Keldysh formalism, the $GW$ self-energy is given by~\cite{Thygesen2007GW,Ness2011GW,2020PonceReview}
\begin{align} \label{eq:sigma_der_GW}
    & \Sigma^{<}_\nk(\omega)
    = \frac{i}{N_q} \sum_{m\mb{q}} \!\int \! \frac{d\omega'}{2\pi} 
    G^{<}_\mkq(\omega+ \omega') W^{>}_\mnk(\mb{q},\omega'),
\end{align}
where $<$ and $>$ denote the lesser and greater components of the corresponding object, respectively.
The imaginary part of the retarded self-energy follows by using the fluctuation-dissipation theorem~\cite{StefanucciBook}:
\begin{equation} \label{eq:sigma_der_Sigma_fdt}
    \mathrm{Im} \Sigma^{\rm R}_\nk(\omega) = -\frac{1}{2f(\omega)} \mathrm{Im} \Sigma^{<}_\nk(\omega),
\end{equation}
where $f(\omega) = 1 / \bigl(e^{\omega / k_{\rm B} T} + 1 \bigr)$ is the Fermi--Dirac occupation function.

The lesser Green function for the electrons reads~\cite{StefanucciBook}
\begin{equation} \label{eq:sigma_der_G<}
    G^{<}_\mkq(\omega) = 2\pi i \delta(\omega - \veps_\mkq) f_\mkq.
\end{equation}
Using the fluctuation-dissipation theorem~\cite{StefanucciBook}, we find the greater component of the electron-electron interaction:
\begin{align} \label{eq:sigma_der_W>}
    &W^{>}_\mnk(\omega)
    = 2i \mathrm{Im} W^{\rm R}_\mnk(\omega) \bigl[1 + n(\omega) \bigr]
    \nnnl
    &= 2i \bigl[ 1 + n(\omega) \bigr] \mathrm{Im}
    \sum_{\beta=0}^{\Nph} \Biggl( 
    \frac{\widetilde{g}^{\rm L}_{mn\beta}(\mb{k},\mb{q}) [\widetilde{g}^{\rm R}_{mn\beta}(\mb{k},\mb{q})]^*}{\omega - \wtw_\qb + i \wtg_\qb}
    \nnnl
    &\qquad-
    \frac{[\widetilde{g}^{\rm L}_{mn\beta}(\mb{k},\mb{q})]^* \widetilde{g}^{\rm R}_{mn\beta}(\mb{k},\mb{q})]}{\omega  + \wtw_\qb+ i \wtg_\qb} \Biggr)
\end{align}

Substituting \Eqs{eq:sigma_der_G<} and \eqref{eq:sigma_der_W>} into \Eq{eq:sigma_der_GW}, we find
\begin{align}
    &\Sigma^{<}_\nk(\omega)
    = \frac{2i}{N_q} \sum_{m\mb{q}} f_\mkq \bigl[ 1 + n(\veps_\mkq - \omega) \bigr] \nnnl
    &\times \mathrm{Im} \sum_{\beta=0}^{\Nph} 
    \Biggl(\frac{\widetilde{g}^{\rm L}_{mn\beta}(\mb{k},\mb{q}) [\widetilde{g}^{\rm R}_{mn\beta}(\mb{k},\mb{q})]^*}{\omega - \veps_\mkq + \wtw_\qb - i \wtg_\qb} 
    \nnnl
    &\qquad\qquad -
    \frac{[\widetilde{g}^{\rm L}_{mn\beta}(\mb{k},\mb{q})]^* \widetilde{g}^{\rm R}_{mn\beta}(\mb{k},\mb{q})]}{\omega - \veps_\mkq - \wtw_\qb - i \wtg_\qb} \Biggr).
\end{align}
Using \Eq{eq:sigma_der_Sigma_fdt} and the identity
\begin{equation}
    \frac{f_\mkq}{f(\omega)} \bigl[ 1 + n(\veps_\mkq - \omega) \bigr]
    = f_\mkq + n(\veps_\mkq - \omega),
\end{equation}
we obtain the imaginary part of the retarded self-energy
\begin{align}
    &\mathrm{Im} \Sigma^{\rm R}_\nk(\omega)
    = -\frac{1}{N_q} \sum_{m\mb{q}} \bigl[ f_\mkq + n(\veps_\mkq - \omega) \bigr] \nnnl
    &\quad\times \mathrm{Im} \sum_{\beta=0}^{\Nph} 
    \Biggl(\frac{\widetilde{g}^{\rm L}_{mn\beta}(\mb{k},\mb{q}) [\widetilde{g}^{\rm R}_{mn\beta}(\mb{k},\mb{q})]^*}{\omega - \veps_\mkq + \wtw_\qb - i \wtg_\qb} 
    \nnnl
    &\qquad\qquad -
    \frac{[\widetilde{g}^{\rm L}_{mn\beta}(\mb{k},\mb{q})]^* \widetilde{g}^{\rm R}_{mn\beta}(\mb{k},\mb{q})]}{\omega - \veps_\mkq - \wtw_\qb - i \wtg_\qb} \Biggr)
    \nnnl
    &= \frac{1}{N_q} \sum_{m\mb{q}} \bigl[ f_\mkq + n(\veps_\mkq - \omega) \bigr]
    \nnnl
    &\quad\times \mathrm{Im} \sum_{\beta=0}^{\Nph} 
    \Biggl(\frac{[\widetilde{g}^{\rm L}_{mn\beta}(\mb{k},\mb{q})]^* \widetilde{g}^{\rm R}_{mn\beta}(\mb{k},\mb{q})}{\omega - \veps_\mkq + \wtw_\qb + i \wtg_\qb} 
    \nnnl
    &\qquad\qquad -
    \frac{\widetilde{g}^{\rm L}_{mn\beta}(\mb{k},\mb{q}) [\widetilde{g}^{\rm R}_{mn\beta}(\mb{k},\mb{q})]^*}{\omega - \veps_\mkq - \wtw_\qb + i \wtg_\qb} \Biggr),
\end{align}
which equals \Eq{eq:el_sigma_plph}.
The real part can be obtained using the Kramers--Kronig relation.

In the zero-broadening limit, we use the relation \Eq{eq:broad_to_nobroad} to find
\begin{align}
    &\mathrm{Im} \Sigma^{\rm R}_\nk(\omega)
    = \frac{1}{N_q} \sum_{m\mb{q}} \bigl[ f_\mkq + n(\veps_\mkq - \omega) \bigr]
    \\
    &\times \mathrm{Im} \sum_{\beta=0}^{\Nph} \sum_{s=\pm}
    \frac{s \abs{\widetilde{g}_{mn\beta}(\mb{k},\mb{q})}^2}{\omega - \veps_\mkq + s\wtw_\qb + i 0^+}
    \nnnl
    &= -\frac{\pi}{N_q} \sum_{m\mb{q}}  \sum_{\beta=0}^{\Nph}
    \abs{\widetilde{g}_{mn\beta}(\mb{k},\mb{q})}^2
    \nnnl
    &\times \sum_{s=\pm} s \bigl[ f_\mkq + n(s\wtw_\qb) \bigr]
    \delta \bigl( \omega - \veps_\mkq + s\wtw_\qb \bigr).
    \nonumber
\end{align}
Using the identity
\begin{align}
    f_\mkq + n(-\wtw_\qb)
    &= f_\mkq - 1 - n(\wtw_\qb) \nnnl
    &= -f^-_\mkq - \wtn_\qb,
\end{align}
we find
\begin{align} \label{eq:ImSigW}
    &\mathrm{Im} \Sigma^{\rm R}_\nk(\omega)
    = -\frac{\pi}{N_q} \sum_{m\mb{q}}  \sum_{\beta=0}^{\Nph}
    \abs{\widetilde{g}_{mn\beta}(\mb{k},\mb{q})}^2
    \\
    &\quad \times \sum_{s=\pm} \bigl[ f^s_\mkq + n(\wtw_\qb) \bigr]
    \delta \bigl( \omega - \veps_\mkq + s\wtw_\qb \bigr),
    \nonumber
\end{align}
which equals \Eq{eq:el_sigma_plph_nobroad}.
The analytical formula for the real part of the self-energy can be obtained from the Kramers--Kronig transformation of Eq.~\eqref{eq:ImSigW}.

\section{Computational details} \label{sec:supp_comp_details}

We performed density functional theory and DFPT calculations using the \textsc{Quantum ESPRESSO} package~\cite{2017GiannozziQE}.
For TiO$_2$, we used an 8$\times$8$\times$8 unshifted $k$-point grid, a kinetic energy cutoff of 80~Ry,
and scalar-relativistic norm-conserving pseudopotentials~\cite{2013HamannONCVPSP} from \textsc{PseudoDojo} (v0.4.1)~\cite{2018VanSettenPseudoDojo} in the Perdew-Wang parametrization of the exchange-correlation functional in the local density approximation~\cite{Perdew1992LDA}.
We used the experimental lattice parameters of $a = 3.784~\mathrm{\AA}$ and $c = 9.515~\mathrm{\AA}$~\cite{Horn1972TiO2} and relaxed the internal coordinates until the forces are below $10^{-5}~\mathrm{Ry/bohr}$.

We used \textsc{Wannier90}~\cite{2020PizziWannier90} and EPW~\cite{Giustino2007,2016PonceEPW} software to generate the Wannier functions and obtain the real-space matrix elements.
We sampled the Brillouin zone with a 4$\times$4$\times$4 grid for both electrons and phonons.
We generated 22 maximally localized Wannier functions without disentanglement using the Ti $d$ and O $p$ orbitals as initial guesses~\cite{1997MarzariMLWF,2012MarzariMLWF}.

We performed Wannier interpolation and computed the dielectric function, plasmon-phonon spectral function, and electron scattering rates using an in-house developed code \texttt{ElectronPhonon.jl} written in the Julia programming language~\cite{2017BezansonJulia}.
Details of the implementation will be discussed elsewhere.
For the calculation of the dielectric function~[\Eq{eq:plph_chi0}], we computed the dielectric function considering states with energy in the range [0~eV, 0.5~eV] with respect to the conduction band minimum energy, using a 50$\times$50$\times$50 $k$-point grid and replace $0^+$ with a smearing of $\eta=15~\mathrm{meV}$ in \Eq{eq:plph_chi0}.
We subtract this broadening from the broadening $\gamma_\qz$ of the plasmon-pole model so that the obtained $\gamma_\qz$ is not sensitive to the value of $\eta$.
We computed the dielectric function for 31 frequency points uniformly sampled from 0 to 500~meV and performed a least-squares fit to compute the plasmon-pole parameters.
To calculate the electron scattering rates~[\Eq{eq:el_sigma_plph}], we summed over the states with energy $\veps_\mkq$ inside the window $[0,\, 0.8~\mathrm{eV}]$ and used a 120$\times$120$\times$120 $q$-point grid and replaced $0^+$ with a smearing of $\eta=10~\mathrm{meV}$ for the bare phonon Green function.
We also applied the same smearing to plasmons if their linewidth $\gamma_\qz$ is below 10~meV.

Calculations for GaAs were performed in a similar manner.
Computational parameters were taken from Ref.~\cite{2021PonceMobility} unless otherwise noted.
We constructed 8 maximally localized Wannier functions~\cite{1997MarzariMLWF,2001SouzaMLWF,2012MarzariMLWF} using sp$_3$ orbitals of Ga and As as initial guesses, with a disentanglement step where the inner (frozen) window spanned up to 3.7~eV above the conduction band minimum energy.
We sampled the Brillouin zone with a 8$\times$8$\times$8 grid for both electrons and phonons.
For the calculation of the plasmon-pole parameters, we used a 200$\times$200$\times$200 $k$-point grid in the energy window [0~eV, 0.3~eV] around the conduction band minimum and a smearing of $\eta=3~\mathrm{meV}$.

\FloatBarrier  

\makeatletter\@input{xx.tex}\makeatother
\bibliography{main}